\documentclass[12pt,preprint]{aastex}

\usepackage{amssymb}
\usepackage{latexsym}
\usepackage{amsmath,graphicx}
\usepackage{epstopdf}

\newcommand{\ueps}{\underline{\epsilon}}

\newcommand{\cntr}{_*}
\newcommand{\dist}{_{\mathrm{dist}}}
\newcommand{\damp}{_{\mathrm{damp}}}
\newcommand{\PLANET}{p}
\newcommand{\planet}{_\PLANET}
\newcommand{\TESTBOD}{t}
\newcommand{\testbod}{_\TESTBOD}
\newcommand{\sA}{\mathcal{S}_A}
\newcommand{\sB}{\mathcal{S}_B}

\newcommand{\De}{\partial_{e}}
\newcommand{\Da}{\partial_{\alpha}}
\newcommand{\Do}{\partial_{\varpi}}
\newcommand{\Dp}{\partial_{\phi}}

\newcommand{\h}{\mathcal{P}}
\newcommand{\hn}{\h_n}
\newcommand{\he}{\h_e}

\newcommand{\R}{\mathcal{R}}

\newcommand{\Q}{\mathcal{Q}}

\newcommand{\s}{\mathcal{S}_1}
\newcommand{\sS}{\mathcal{S}_2}
\newcommand{\Rd}{\mathcal{R}_D}

\newcommand{\fss}{f_{s,1}}
\newcommand{\fsS}{f_{s,2}}

\newcommand{\pomdiff}{\left(\Delta \varpi \right)}

\newcommand{\FI}{\mathcal{F}_{\mathrm{I}}}
\newcommand{\FE}{\mathcal{F}_{\mathrm{E}}}

\newcommand{\J}{\mathcal{J}}
\newcommand{\kk}{\mathcal{K}}
\newcommand{\LL}{\tilde \J}
\newcommand{\N}{\mathcal{N}}
\newcommand{\tN}{\tilde\N}
\newcommand{\E}{\mathcal{E}}
\newcommand{\Ew}{\E_\varpi}
\newcommand{\Ep}{\E_\phi}
\newcommand{\tE}{\tilde\E}
\newcommand{\C}{\mathcal{C}}
\newcommand{\G}{\mathcal{G}}

\newcommand{\eh}{h_e}
\newcommand{\ah}{h_\alpha}

\newcommand{\be}{\begin{equation}}
\newcommand{\ee}{\end{equation}}
\newcommand{\ba}{\begin{aligned}}
\newcommand{\ea}{\end{aligned}}

\begin{document}

\title{\bf MEAN MOTION RESONANCES IN EXOPLANET SYSTEMS:\\ 
AN INVESTIGATION INTO NODDING BEHAVIOR }

\author{Jacob A. Ketchum,$^{1}$, Fred C. Adams,$^{1,2}$ 
and Anthony M. Bloch$^{1,3}$} 

\affil{$^1$Michigan Center for Theoretical Physics \\
Physics Department, University of Michigan, Ann Arbor, MI 48109} 

\affil{$^2$Astronomy Department, University of Michigan, Ann Arbor, MI 48109} 

\affil{$^3$Department of Mathematics, University of Michigan, 
Ann Arbor, MI 48109} 

\begin{abstract}

Motivated by the large number of extrasolar planetary systems that are
near mean motion resonances, this paper explores a related type of
dynamical behavior known as ``nodding''. Here, the resonance angle of
a planetary system executes libration (oscillatory motion) for several
cycles, circulates for one or more cycles, and then enters once again
into libration. This type of complicated dynamics can affect our
interpretation of observed planetary systems that are in or near mean
motion resonance.  This work shows that planetary systems in (near)
mean motion resonance can exhibit nodding behavior, and outlines the
portion of parameter space where it occurs. This problem is addressed
using both full numerical integrations of the planetary systems and
via model equations obtained through expansions of the disturbing
function. In the latter approach, we identify the relevant terms that
allow for nodding. The two approaches are in agreement, and show that
nodding often occurs when a small body is in an external mean motion
resonance with a larger planet. As a result, the nodding phenomenon
can be important for interpreting observations of transit timing
variations, where the existence of smaller bodies is inferred through
their effects on larger, observed transiting planets. For example, in
actively nodding planetary systems, both the amplitude and frequency
of the transit timing variations depend on the observational time
window.

\end{abstract}


\section{Introduction} 

The current observational sample of extrasolar planets includes many
systems with multiple planets, and many systems have orbital period
ratios that are close to integer values (e.g., Fabrycky et al. 2012).
These systems are thus candidates for being in mean motion resonance
(MMR), which represents a special dynamical state for a planetary
system. In addition to the necessary period ratio, the other dynamical
variables of a resonant system must allow one or more resonance angles
(see below for their definitions) to execute oscillatory behavior
(e.g., Murray \& Dermott 1999; hereafter MD99).  One way to describe
this requirement is that the resonant angle(s) must reside in a
``bound state'' within an ``effective potential well''.  Because of
the special conditions required for a planetary system to reside in
mean motion resonance, systems found in such states must have a
constrained dynamical history. The relative fraction of planetary
systems in mean motion resonance thus provides important information
regarding planetary formation and early dynamical evolution.

The dynamics of mean motion resonances is often more complex than
indicated by standard textbook treatments.  This paper explores one
such complication called ``nodding'', where the resonance angle
librates for several cycles and then circulates for one or more cycles
before returning to an oscillatory state. This paper addresses the
problem using two complementary approaches, i.e., both full numerical
integrations of the planetary systems and the construction of model
equations obtained through expansions of the disturbing function. The
expansion approach produces a large number of terms, and we identify
the ones that allow for nodding behavior. These two approaches are in
good agreement, and show that nodding often occurs when a small body
is in mean motion resonance with a larger planet. The immediate goal
of this paper is to obtain a better understanding of nodding in the
context of three-body planetary systems. The over-arching goal is to
provide a more detailed basis for interpreting observed systems that
are found in or near resonance, including systems that exhibit transit
timing variations (TTVs), which provide a means of detecting small
bodies interacting with larger planets in transit (Agol et al. 2005).

We note that previous dynamical studies, especially concerning
asteroids in our Solar System, have found behavior that is
qualitatively similar to the nodding phenomenon explored herein. For
example, asteroids near the 2:1 MMR with Jupiter are predicted to
alternate between two modes of libration (Greenberg \& Franklin 1975).
In one mode, the longitude of perihelion for the asteroid librates
about the longitude of perihelion for Jupiter; in the other mode, the aphelion of the asteroid
librates about the longitude of conjunction.  The asteroids are
predicted to alternate between the two different modes. Similar
apocentric resonances are found for the Hildas asteroid group (see
Henrard et al. 1986; Ferraz-Mello 1988). The phase space of these
dynamical systems contain separatrices for both global and internal
(secondary) resonances (Morbidelli \& Moons 1993); chaotic motion near
the separatrices can lead to crossing, and hence to alternating modes
of libration (see also Michtchenko et al. 2008ab).  In addition to
appearing in planetary systems, this nodding behavior arises in other
dynamical systems, including the driven, inverted pendulum (Acheson
1995).  Additionally, nodding systems sometimes exhibit similar phase
characteristics to other known dynamical systems, such as the the
Duffing oscillator (Guckenheimer \& Holmes 1983).

Both nodding and transit timing variations can occur in planetary
systems in or near mean motion resonance. Observed resonant and
near-resonant systems provide important information about planetary
systems.  As one example, external perturbations can remove systems
from resonance if the perturbations are large enough and if they act
over a sufficiently long time (Adams et al. 2008; Rein \& Papaloizou
2009; Lecoanet et al. 2009; Ketchum et al. 2011a); furthermore, such
conditions can be realized in the circumstellar disks that form
planets. As a result, planetary systems that are observed in resonance
today must not have been greatly perturbed in the past, or they must
have been subsequently influenced by significant dissipative
interactions. On the other hand, resonances can also act as a
protection mechanism, especially for close and massive planets.
Indeed, the current sample of exoplanets includes systems that can
only exist because of strong resonant interactions; the GJ876 system
provides one such example (see Snellgrove et al. 2001; Beauge \&
Michtchenko 2003; Kley et al. 2005).

As another example, we note that entry into mean motion resonance is
non-trivial: If the orbital elements of a planetary system are
selected at random, the chance that the system resides in a mean
motion resonance is relatively modest (even when the period ratio is
chosen to be near the ratio of small integers). However, systems can
evolve into resonance states through the process of convergent
migration (e.g., Lee \& Peale 2002), where, for instance, the outer
planet migrates inward faster than the inner planet, and the two
bodies subsequently move inward together. Even in this scenario,
survival of the resonance can be compromised by overly rapid migration
(Quillen 2006), and/or by turbulent forcing from the disk driving the
migration (Lecoanet et al. 2009; Ketchum et al. 2011a).

As one potential application of this work, nodding can affect our
interpretation of TTVs (Agol et al. 2005). In this setting, unseen
small bodies orbit outside observed transiting planets (usually Hot
Jupiters). The smaller bodies affect the orbit of the inner, larger
planets and lead to small variations in the timing of the transit
events. This phenomenon is potentially a powerful method to detect
(infer the presence of) smaller, otherwise unobserved, planets in such
systems. Indeed, discoveries of this type have already been reported
(e.g., Holman et al. 2010; Cochran et al.  2011), and many more are
expected in the near future.  However, the timing variations are
largest when the smaller planets are in or near mean motion resonance
with the larger planet (e.g., Nesvorn{\'y} \& Morbidelli 2008), and
such systems are susceptible to nodding as studied herein. Even
without the complication posed by nodding, inferring the system
properties from observations is a sensitive process (e.g., Veras et
al. 2011).  In any case, the results of this work will be useful for
future interpretation of systems that exhibit transit timing
variations.

This paper is organized as follows. In Section \ref{S:full}, we study
the nodding phenomena through numerical integrations of multiple
planet systems that are near mean motion resonance. This investigation
shows that complex dynamical behavior, including nodding, is often
present, and outlines the portion of parameter space where it
occurs. For systems that exhibit nodding behavior, we then outline the
corresponding effects on transit timing variations.  In Section
\ref{S:eom}, we derive a class of model equations to describe nodding
behavior.  Here we expand the disturbing function for planetary
interactions (e.g., MD99), keep the highest order terms, and identify
the relevant terms that lead to nodding. The resulting model equations
elucidate the dynamical ingredients required for nodding behavior to
take place.  We conclude, in Section 4, with a summary of our results,
a discussion of their implications, and a brief description of future
work.

\section{Numerical Study of Nodding}\label{S:full}

\subsection{Full 3-body Numerical Simulations}

This paper studies nodding of mean motion resonance angles for planet
pairs that are near MMR. For simplicity, most of this work focuses on
the 2:1 resonance.  The term ``nodding'' here refers to a tendency to
repeat a pattern of bounded libration for several cycles followed by
one or more cycles of circulation.  For intermediate times, the system
exhibits behavior of MMR, but intermittent bouts of circulation may
produce a cumulative net circulation of the resonance angle over many
libration times.  In the context of resonant angle phase trajectories,
nodding can be described as motion near a separatrix in the phase
space. The phase space for internal resonances contain one separatrix,
whereas the phase space for external resonances can contain two
distinct separatrices.  The qualitative differences occur, in part,
due to the existence of asymmetric external resonances, which arise
when the orbital eccentricity for the test body becomes sufficiently
large (Ferraz-Mello et al. 2003). The existence of asymmetric
resonance strongly depends on the mass ratio (see Michtchenko et al
2008b); for example, for $m_2 \ll m_1$, all stable orbits are
asymmetric (Beauge 1994).

To carry out the study in this section, we numerically integrate the
three-body gravitational forces using a Bulirsch-Stoer integration
scheme.  In addition to gravity, both general relativistic corrections
and stellar tidal damping are included in force calculations, but the
inclusion of these additional forces (which are small) are not
necessary to produce the interesting features explored in this work.
Our system consists of a star with mass $M_{*} = 1 M_{\sun}$, a
massive planet ($m\planet = 1 M_{jup}$), and a test body ($m = 10^{-6}
M_{jup}$).  We chose this particular test mass in order to minimize
its influence on the planet's motion and to obtain the clearest
dynamical signature of resonance angle nodding. However, nodding is
also present for larger masses for the third body (in Section 2.4 we
consider masses as large as $m=10 M_\earth$).  The Jovian planet is
placed in orbit with period $T\planet = 1$ year, and the test body is
placed in orbit with initial period $T = 1/2$ or $T = 2$ years for
studies involving internal and external resonances, respectively. We
choose a benchmark test body eccentricity of $e = 0.15$, which is
motivated by previous work (Ketchum et al. 2011a), and we choose from
two values for the planet's initial orbital eccentricity, $e\planet$ =
0.001 or $e\planet$ = 0.1.

To fully describe the initial configuration of the system, the initial
orbital angles must be specified. We parameterize this study using the
set of angles given by $\Delta \varpi_0$ -- the relative alignment
between the two orbits -- and $f_0$ -- the test body's true anomaly.
Unless noted otherwise, all simulations begin with the planet and test
body in conjunction.  In an attempt to sample the available resonance
angle phase space resulting from a choice of initial orbital elements
$\{ e\planet, e, \alpha, \Delta \varpi \}$, values of $f_0$ spanning
the full range $-\pi$ to $\pi$ in increments of $\pi/100$ are
sampled. Following this systematic approach, those states that have
phase trajectories occurring near a separatrix, which ultimately lead
to nodding, are easily found.

The simulations are integrated for $10^4$ years, a sufficient amount
of time to capture several secular cycles for most initial
configurations (for completeness, note that the parameter space for
external resonances contain small regions where the secular cycle's
period is infinite -- see Michtchenko et al. 2008b). The energy for a
typical system that experiences no significant close encounters is
conserved to better than one part in $10^{10}$. The planets' period
ratio is monitored to confirm that the system remains nearly integer
commensurate during the integration and hence that near resonance has
not been compromised by a chance close encounter. The osculating
elements for both bodies are recorded once per orbit of the inner
planet. Since the libration timescales are of order $\sim30-100$
orbits (albeit with large variations), this sampling cadence is
frequent enough to resolve the behavior of the resonance angles.  
For 2:1 resonances, the resonance angles of interest are 
\be\label{E:phi_int}
\phi = 2\lambda\planet - \lambda - \varpi \ ,
\ee
for a test body internal to the planet's orbit, and
\be\label{E:phi_ext}
\phi = 2\lambda - \lambda\planet - \varpi \ ,
\ee
for a test body external to the planet's orbit, where $\lambda$ is the
mean longitude and subscript $\PLANET$ denotes the orbital elements
belonging to the Jovian planet. And finally, the time rate of change
for the resonance angle, $\dot{\phi}$, is determined by quadratic
interpolation, and used to construct resonance angle phase
trajectories.

\subsection{Nodding Features for Near-Resonance}

The range of dynamical behaviors encountered in (or near) mean motion
resonance is surprisingly rich (e.g., Michtchenko et al. 2008ab).
Presented here is a small representative sample set of the simulations
outlined above which display the main features found in these
resonance states.

\begin{figure}[htbp]
\begin{center}
\includegraphics[width = 140mm]{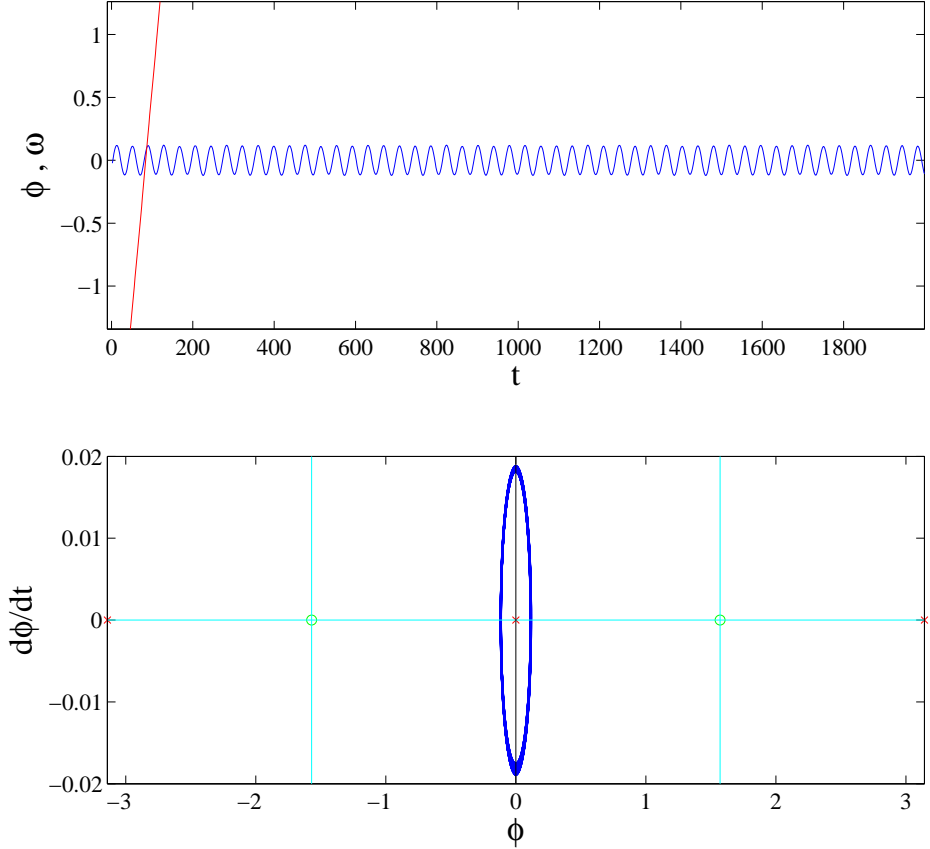}
\caption{Internal 2:1 near-resonance with periaspses initially
anti-aligned ($\Delta \varpi = \pi$), test body near periastron at
conjunction ($f_0 \approx 0$), and $e\planet = 10^{-3}$. All angles
are in radians and time is in years.  The \emph{top panel} shows the
resonance angle, $\phi$ (blue curve), and the angle of apsides,
$\Delta \varpi$ (red curve). In this example, resonance angle nodding
does \emph{not} occur.  The \emph{bottom panel} shows the phase
trajectory of the resonance angle during the full $10^4$ years of
simulation. This system is close to the text book example of mean
motion resonance described by the pendulum model (see Murray \&
Dermott 1999).  The libration amplitude is small ($\Delta \phi \sim
0.1$ rad) and the equilibrium point for oscillations is at $\phi =
0$. }
\label{F:Int_anti}
\end{center}
\end{figure}

We first consider systems where nodding does \emph{not} occur (see
Figure \ref{F:Int_anti}). The figure features a system with a
configuration and behavior deviating only slightly from the pendulum
model of MD99. For this system, the perturbing planet is placed in
orbit with semi-major axis $a\planet=1$AU and eccentricity $e\planet =
10^{-3}$ around a $1M_\sun$ star.  A test particle of mass $m =
10^{-6} M_{jup}$ is set in a coplanar orbit with semi-major axis $a =
0.63$ AU and orbital eccentricity $e = 0.15$, which places the two
orbiting bodies near a 2:1 period ratio.  The orbits are initially
anti-aligned ($\Delta \varpi_0 = \pi$) and the orbiting bodies placed
in conjunction with the test particle near periapse -- the system is
prepared such that the planet's influence on the test body's motion is
initially minimized, i.e., the two orbiting bodies cannot be separated
further from one another during conjunction given this set of orbital
elements.  The top panel of Figure \ref{F:Int_anti} shows the
resonance angle, $\phi$, from equation (\ref{E:phi_int}) in blue and
the angle of apsides, $\Delta \varpi$, in red. The resonance angle
librates with a small amplitude of $\Delta \phi \simeq 0.1$ radians
around the equilibrium $\phi = 0$, while the apsidal angle circulates
due to a prograde motion of the test particle's longitude of
periastron -- the planet's longitude of periastron, $\varpi\planet$,
does not move significantly. In the limit of small orbital
eccentricities $e$ and $e\planet$, the test body's true anomaly
approximately coincides with the resonance angle at instances of
conjunction. Thus, for this system, these dynamics depend mainly upon
our choice for $|f_0|$ and are independent of our choice for $\Delta
\varpi$. The bottom panel of Figure~\ref{F:Int_anti} shows the phase
trajectory for the resonance angle. This panel depicts a phase
trajectory analogous to small oscillations of a simple pendulum, as
expected. In this respect, the pendulum model of the circular
restricted three body problem provides a sufficient model for
resonances arising from orbital configurations of this type --
although the planet's orbital eccentricity is non-zero, dynamical
deviations from the pendulum model of the circular restricted three
body problem incurred through small departures from circular symmetry
are negligible.

\begin{figure}[htbp]
\begin{center}
\includegraphics[width = 150mm]{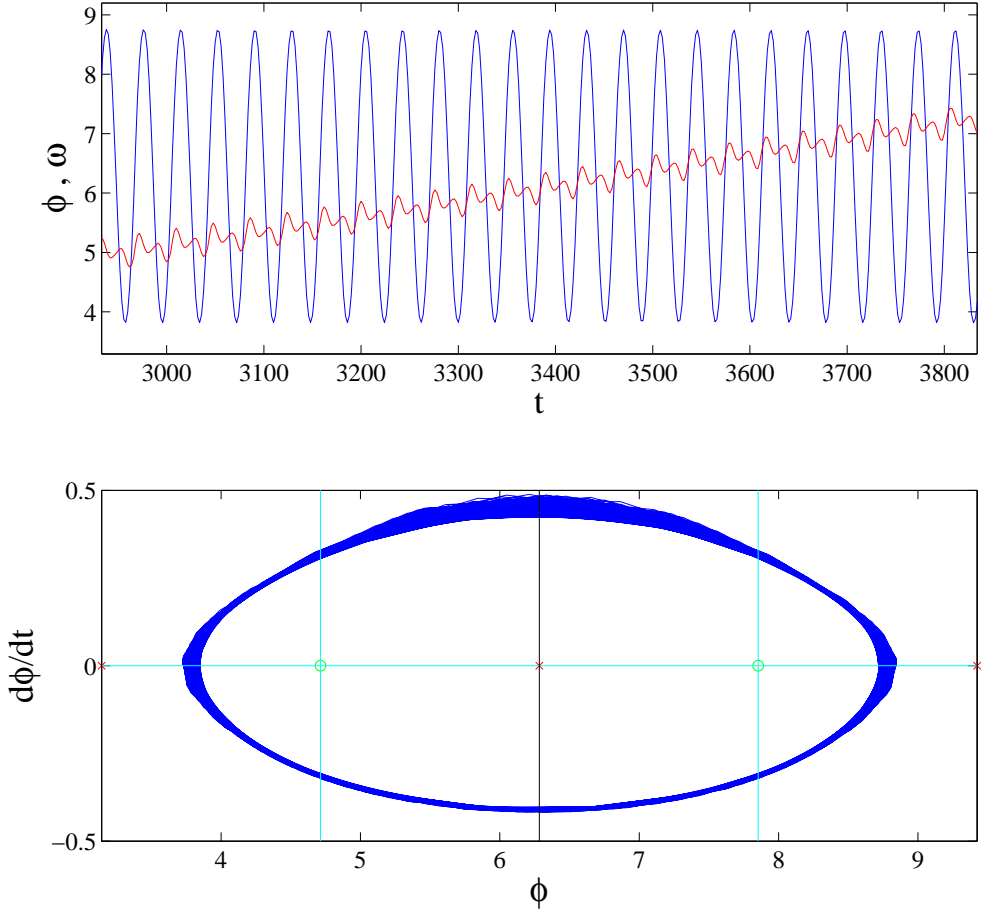}
\caption{Internal 2:1 near-resonance with periapses initially  
anti-aligned ($\Delta \varpi = \pi$), test body closer to apastron than
to periastron during initial conjunction with a true anomaly $f_0
\approx 2\pi/3$, and the perturbing planet's orbit nearly circular
with an eccentricity of $e\planet = 10^{-3}$.  All angles are in
radians and time is in years.  The \emph{top panel} depicts the
resonance angle (blue curve) and the angle of apsides (red curve).
The resonance angle undergoes large amplitude oscillations of $\Delta
\phi \ge 3 \pi /4$ due to the test body's large angular displacement
from periastron of its orbit during the initial conjunction with the
perturbing planet. In this example, resonance angle nodding does
\emph{not} occur. The \emph{bottom panel} shows the phase trajectory
for the resonance angle over the full $10^4$ years of the simulation,
which oscillates about 0 (modulo $2\pi$) and doesn't circulate at
anytime.}
\label{F:Int_align}
\end{center}
\end{figure}

As $|f_0|$ increases from 0 to $\pi$, the amplitude of oscillations
also increase until the conjunction line approaches apoapse, where the
system reaches a separatrix and the resonance angle will circulate
rather than oscillate.  Figure~\ref{F:Int_align} shows a system
identical to that of Figure~\ref{F:Int_anti} in all aspects except
that $f_0 \approx 2\pi/3$ radians, so that the resonance angle
exhibits larger oscillations. The libration frequencies of the two
systems however, are practically identical with frequency $\omega \sim
0.03 \text{ yr}^{-1}$.

\begin{figure}[htbp]
\begin{center}
\includegraphics[width = 100mm]{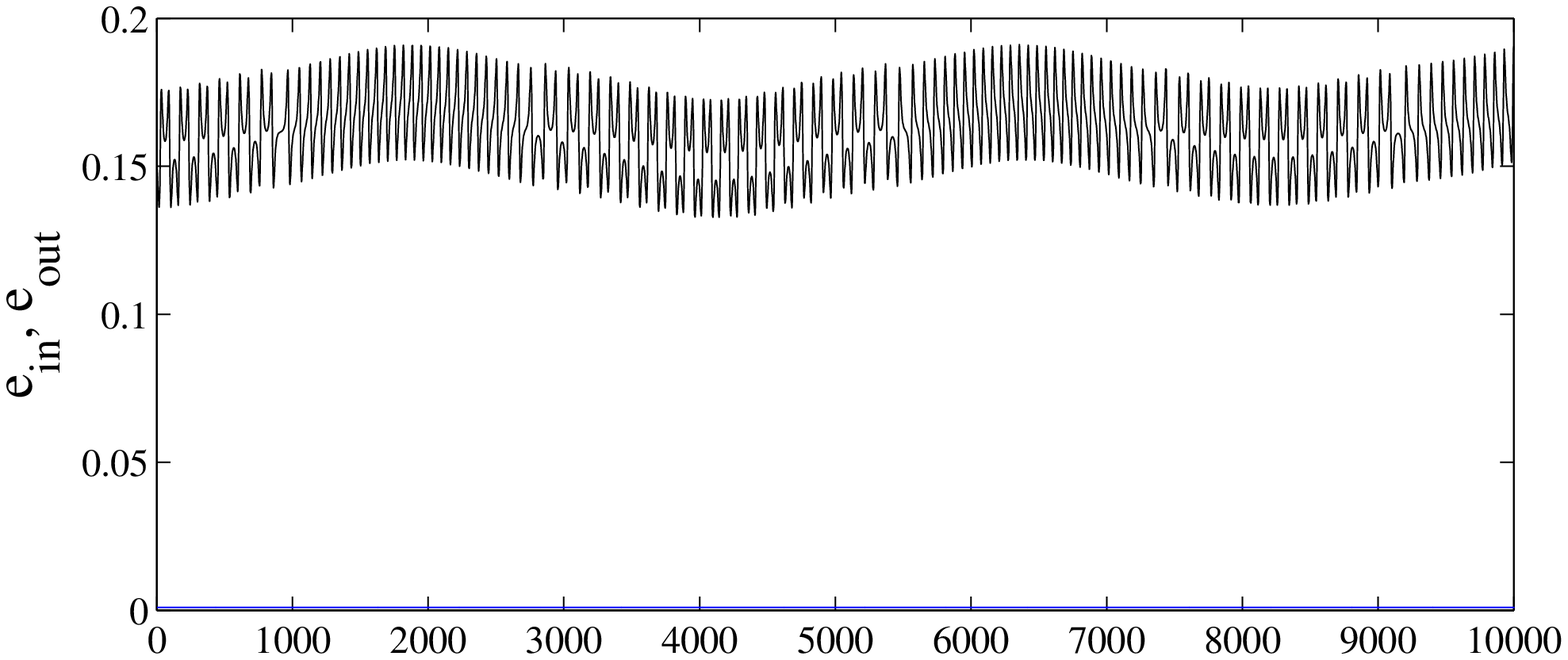}
\includegraphics[width = 100mm]{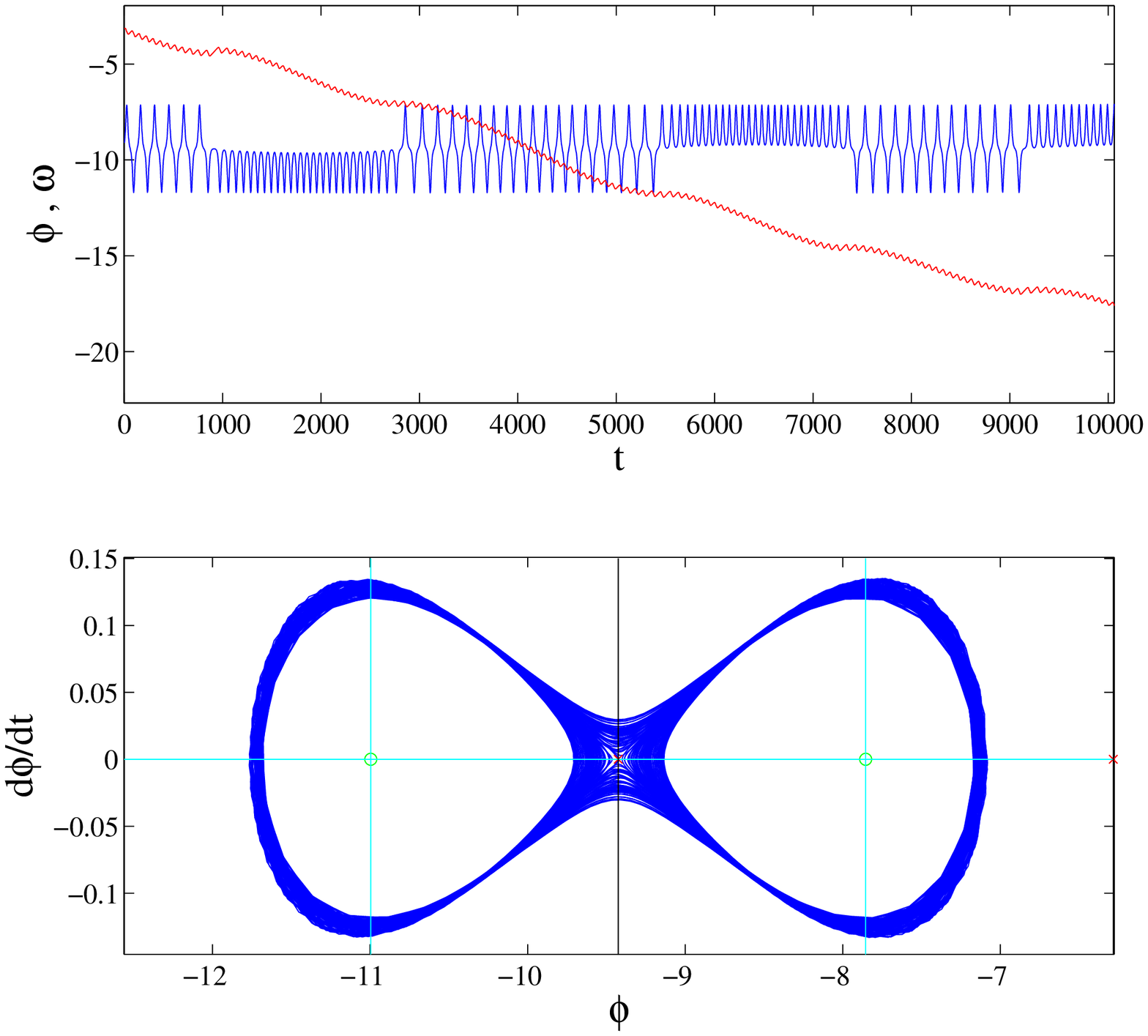}
\caption{External 2:1 near-resonance with periapses initially  
anti-aligned ($\Delta \varpi \approx \pi$), test body near apastron at
conjunction ($f_0 \approx \pi$), and a nearly circular planetary orbit
with eccentricity $e\planet = 10^{-3}$.  All angles are in radians and
time is in years.  The \emph{top panel} shows the time series of the
test body's eccentricity (black curve -- taking values between $0.14
\lesssim e \lesssim 0.18$) and the planet's eccentricity (blue curve
-- nearly constant at $e\planet \simeq 10^{-3}$). 
The \emph{middle panel} shows the resonance angle, $\phi$ (blue
curve), and the angle of apsides, $\Delta \varpi$ (red curve).  The
resonance angle jumps between two points of attraction located near
$\pi/2$ and $3 \pi / 2$ (mod $2 \pi$) -- sometimes librating around
one of these two points many times (e.g., for times between $1000
\lesssim t \lesssim 3000$) before jumping to the opposite point of
attraction -- this is one form of nodding.  The \emph{bottom panel}
shows the resonance angle's phase trajectory for the entire $10^4$
years of simulation.  This trajectory traces out a complex dance
around the inner separatrix of the phase space of the external 2:1
asymmetric resonance.  } 
\label{F:Ext_anti}
\end{center}
\end{figure}

For external resonance scenarios, the behavior is significantly
different. Figure \ref{F:Ext_anti} shows the first example of
nodding. In the figure, $|f_0| \sim \pi$ so initial conjunction is
close to apastron of the test body's orbit, and the orbits are
initially anti-aligned, $\Delta \varpi_0 = \pi$. This set of initial
orbital elements gives the maximum possible spatial separation between
the planet and test body. The resonance angle $\phi$, given by
equation (\ref{E:phi_ext}) and shown in blue in the top panel of the
figure, seems to be attracted to one of two stable fixed points, with
an unstable fixed point effectively located at $\phi = \pi$.  This is
an example of an asymmetric resonance (Lee 2004; compare with
Callegari et al. 2004), and the existence of the two equilibrium
points on either side of $\phi = \pi$ are due to a bifurcation
occurring in the dynamics for sufficiently large test body orbital
eccentricity (e.g., Michtchenko et al. 2008b).  In this case, the
system was configured near a different kind of separatrix than those
encountered for internal resonances, and motion near this separatrix
leads to nodding behavior that is unique to the external case. In
general, the planar three-body problem has two degrees of freedom,
with two proper frequencies: resonant and secular ones. As a result,
oscillations of the problem (including nodding) must be combinations
of the two proper frequencies.

\begin{figure}[htbp]
\begin{center}
\includegraphics[width = 150mm]{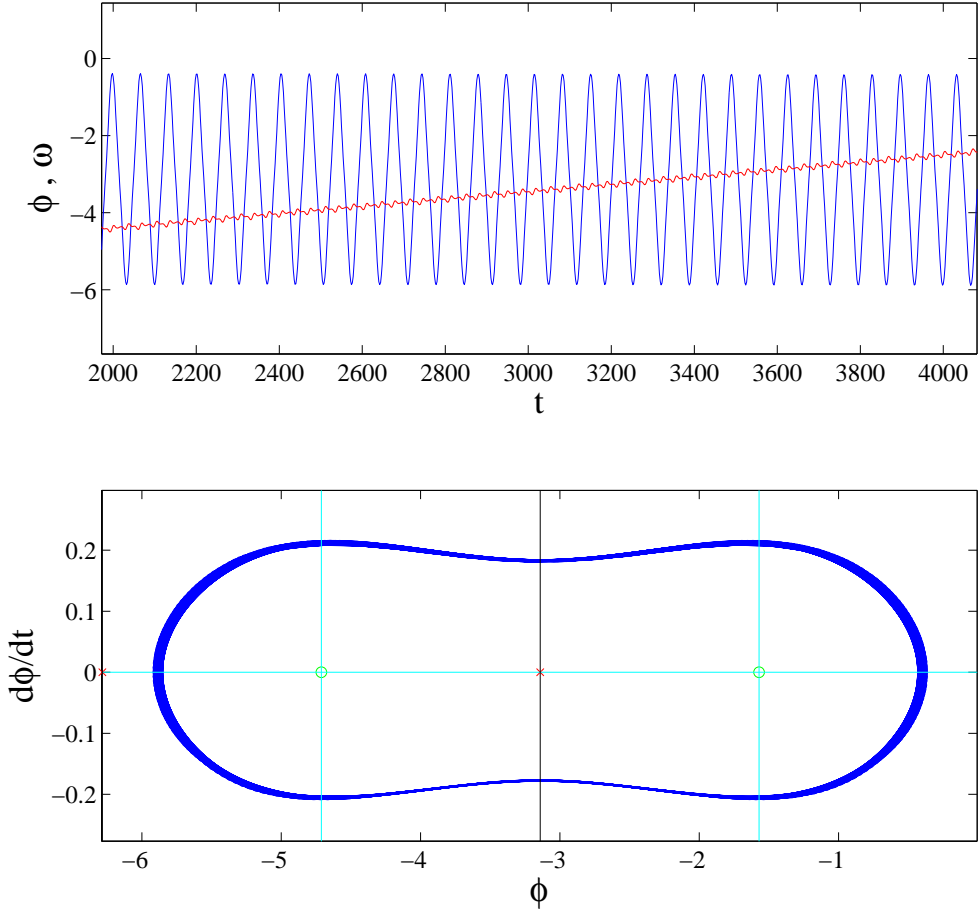}
\caption{External 2:1 near-resonance with periapses initially 
anti-aligned ($\Delta \varpi \approx \pi$), test body displaced
$\approx 40^{\circ}$ from periastron at conjunction ($f_0 \approx
\pi/4$), and a nearly circular planetary orbit with eccentricity
$e\planet = 10^{-3}$. All angles are in radians and time is in years.
The \emph{top panel} shows the resonance angle, $\phi$ (blue curve),
and the angle of apsides, $\Delta \varpi$ (red curve).  The resonance
angle repeatedly jumps between two points of attraction located near
$\pi/2$ and $3 \pi / 2$ (mod $2 \pi$) -- this behavior is one form of
nodding.  The \emph{bottom panel} shows the resonance angle's phase
trajectory for the entire $10^4$ years of simulation. The resonance
angle's motion temporarily slows as it passes apoastron of the test
body's orbit $ \phi \approx -\pi$, and accelerates producing sharper
peaks in the resonance angle signal of the top panel, in contrast to
the internal resonance case (compare with Figure \ref{F:Int_align}).}
\label{F:Ext_align}
\end{center}
\end{figure}

Note that there are several secular cycles shown in Figure
\ref{F:Ext_anti}, and during each secular cycle the resonance angle
seems to choose one of three different libration modes -- \emph{(a)}
resonance oscillations enclosing some point to the left of $\phi =
\pi$, \emph{(b)} oscillations enclosing some point to the right of
$\phi = \pi$, or \emph{(c)} oscillations enclosing all three
points. As $|f_0| \rightarrow 0$ for similarly prepared systems,
resonant oscillation amplitudes reach the maximum possible value
$\Delta \phi \approx \pi$ radians before circulating.  Encountered
here is another separatrix, which is analogous to the separatrix for
the internal resonance. In contrast to the internal resonance
separatrix, the phase space trajectory for large amplitude
oscillations here takes a different shape -- the resonant angle's
speed decreases (increases) as it approaches (departs from) the test
body's apoapse location. Systems on such phase trajectories appear to
nod once per one resonant libration, as depicted in Figure
\ref{F:Ext_align} where $f_0 \approx \pi/4$.  In this regard, the
resonance angle moves as if it lives in a quartic potential, with a
local maximum at $\phi = \pi$, two minima near $\phi = \pi/2$ and
$\phi = 3\pi/2$, and a maximum at $\phi = 0$. This feature of the
motion near the outer separatrix of an asymmetric outer resonance can
lead to a period increase or decrease by a factor of two for transit
timing variations, a result presented later on in this paper.

\begin{figure}[htbp]
\begin{center}
\includegraphics[width=100 mm]{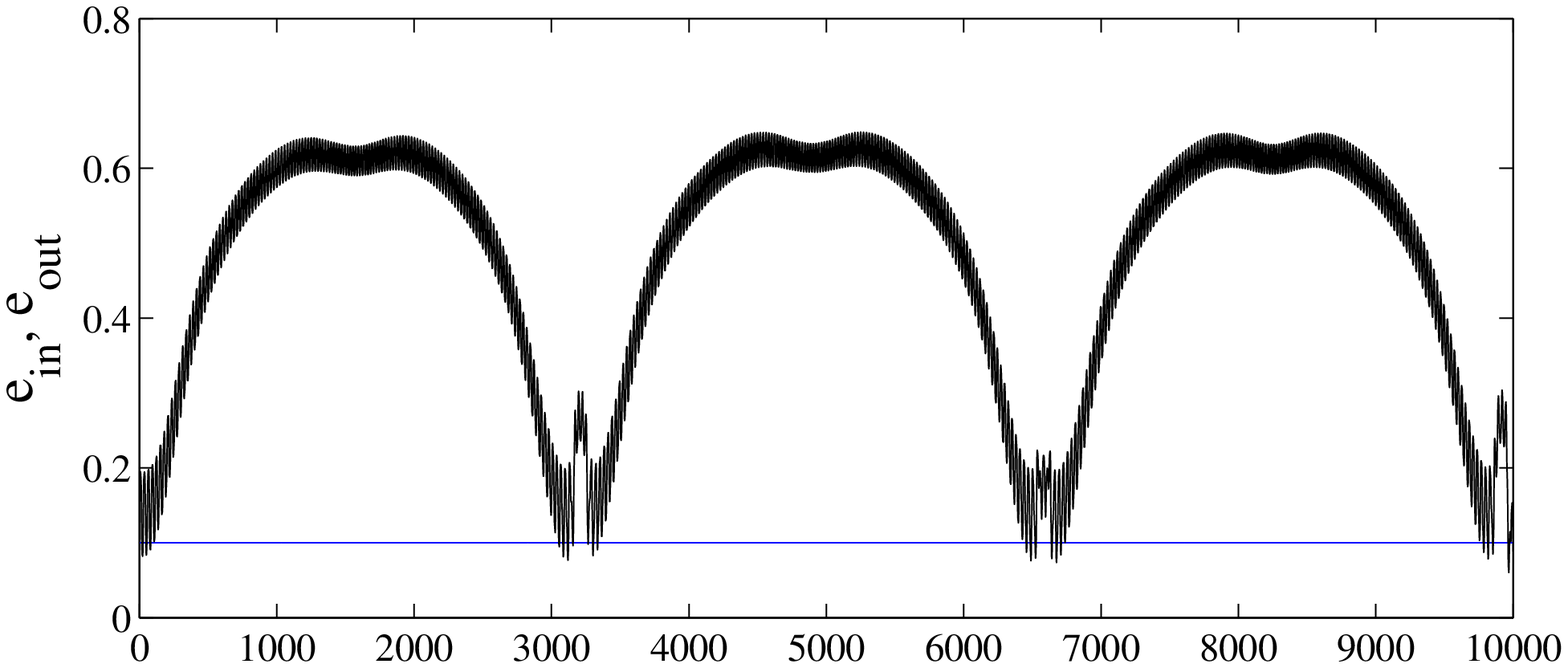}
\includegraphics[width=100 mm]{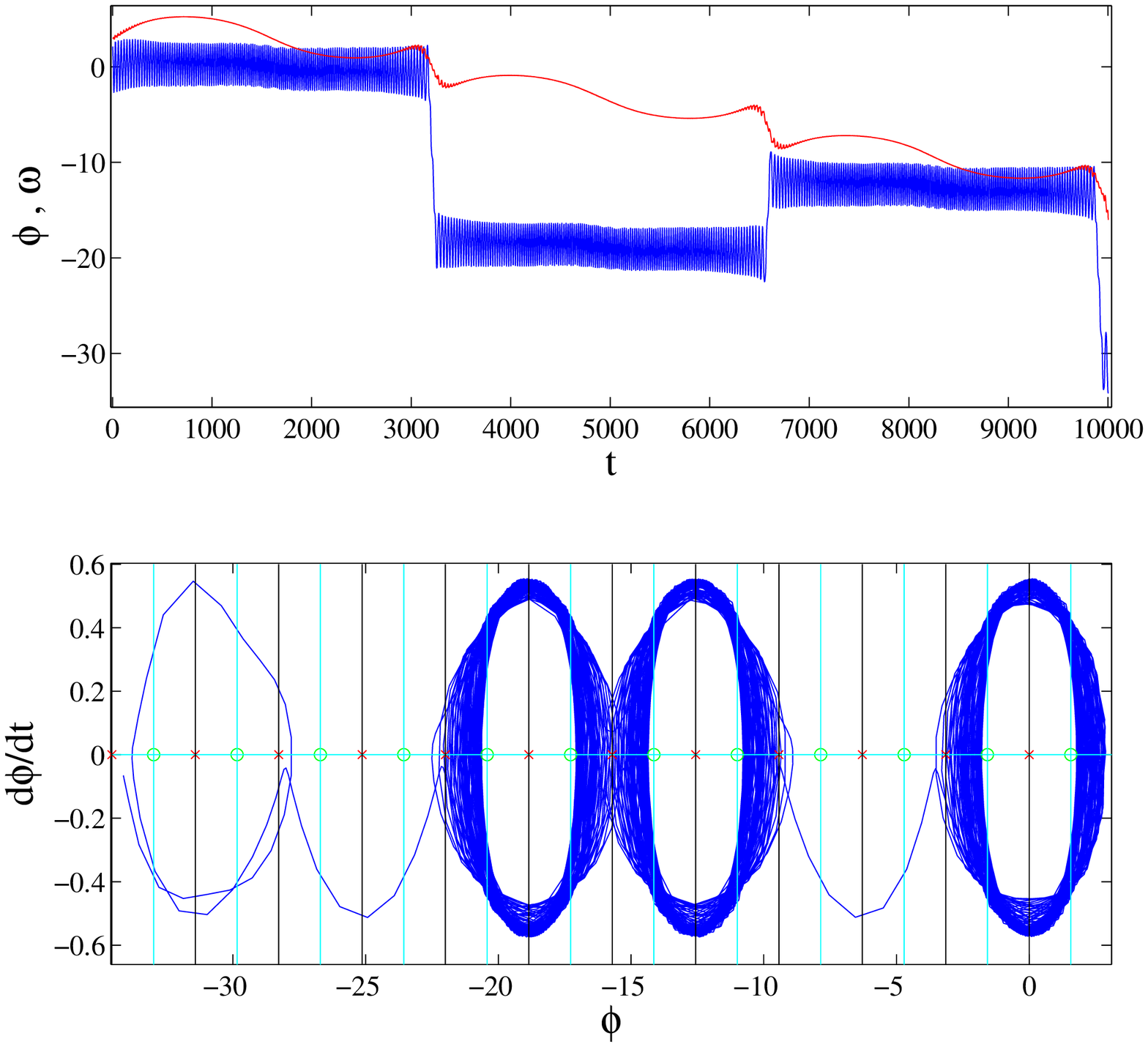}
\caption{Internal 2:1 near-resonance, periapses initially aligned
($\Delta \varpi \approx 0$), test body near apastron during initial
conjunction ($f_0 \approx \pi$), and the external planet's orbit with
eccentricity is $e\planet = 0.1$.  All angles are in radians and time
is in years.  The \emph{top panel} shows the eccentricity time series
for both the test particle (black curve) and the planet (blue curve).
The test body's eccentricity reaches values in excess of $e \gtrsim
0.6$ and as low as $e \lesssim 0.1$, the planet's orbital eccentricity
does not vary significantly. The \emph{middle panel} shows the time
series for the resonance angle (in blue) and $\Delta \varpi$ (in red).
The resonance angle librates about $\phi \approx 0$ with a varying
amplitude $ \pi/2 \lesssim \Delta \phi \lesssim \pi$ for large
stretches of time, and suddenly circulates once (at time close to $t
\approx 6500$ years) or twice (at time $t \approx 3000$ years). This
is a form of nodding. The moments in time where the nodding events
occur are correlated with the times that the test body's orbital
eccentricity is comparable to the planet's orbital eccentricity, $e
\simeq e\planet$, and to times when the orbits are aligned (the angle
of apsides $\Delta \varpi \approx 0$ modulo $2\pi$). The 
\emph{bottom panel} shows the corresponding phase trajectory of the
resonance angle, $\dot{\phi}$ vs $\phi$.}
\label{F:Int_nod}
\end{center}
\end{figure}

\begin{figure}[htbp]
\begin{center}
\includegraphics[width=120 mm]{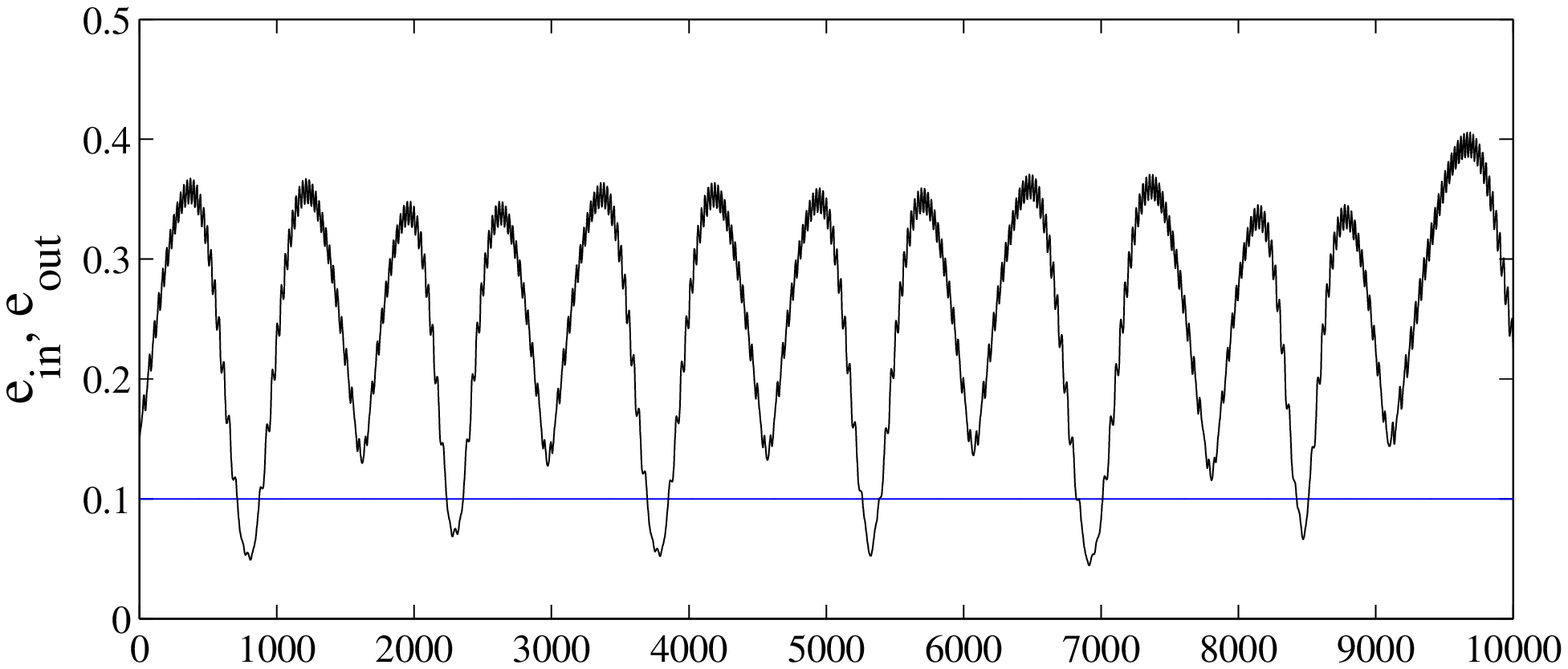}
\includegraphics[width=120 mm]{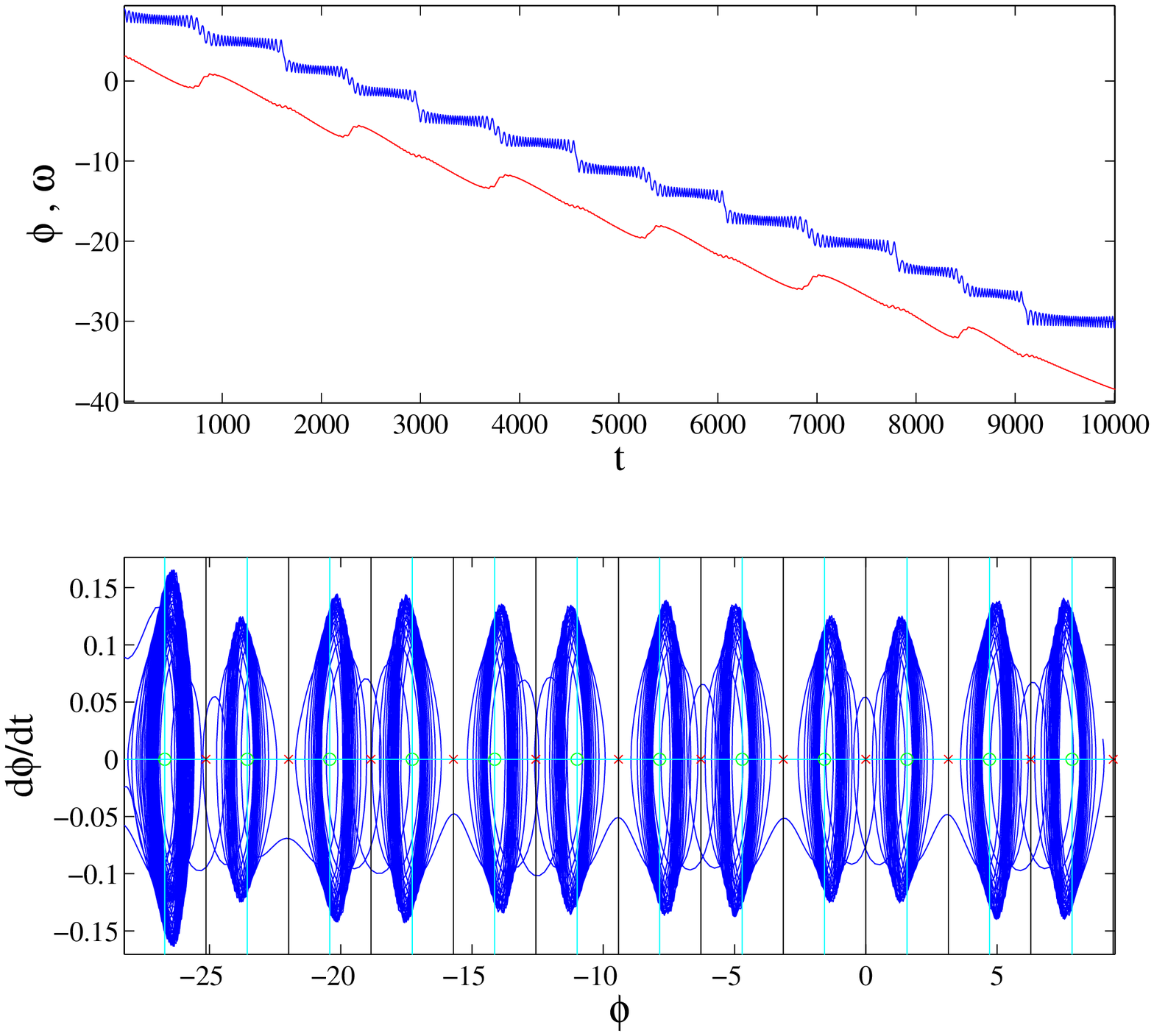}
\caption{External 2:1 near-resonance, periapses anti-aligned 
($\Delta \varpi \approx \pi$), test body near apoastron during initial  
conjunction ($f_0 \approx \pi$), and planet on an eccentric orbit with
$e\planet = 0.1$. All angles are in radians and time is in years.  The
\emph{top panel} shows the eccentricity time series for both the test
particle (black curve) and the perturber (blue curve).  The planet's
eccentricity remains nearly constant during the simulation, while the
test body's eccentricity reaches values up to $e \approx 0.4$ and down
to $e \approx 0.05$. The \emph{middle panel} shows the time series for
the resonance angle (in blue) and $\Delta \varpi$ (in red).  The
\emph{bottom panel} shows the phase trajectories of the resonance
angle, $\dot{\phi}$ vs $\phi$. } 
\label{F:Ext_nod}
\end{center}
\end{figure}

The nodding features become more prevalent as the perturber's
eccentricity increases, where the resonance angle can begin
accumulating a net circulation over longer times.  Figures
\ref{F:Int_nod} and \ref{F:Ext_nod} provide examples of the type of
circulation behavior we observe in simulations. These two figures also
showcase the stark contrasts between nodding for internal and external
resonances, respectively. In this comparison, the planet's orbital
eccentricity is $e\planet = 0.1$ -- substantial enough to be well
outside the \emph{circular} restricted three body regime (where the
pendulum model is strictly valid). Both examples show the tendency for
$\Delta \varpi$ to circulate, on average, with a periodic component
resulting from secular interactions, and, on average, the resonance
angle circulates at the same rate. For times shorter than secular
timescales, however, the resonance angle librates with some amplitude,
not exceeding $\Delta \phi = \pi$, around some equilibrium point. The
point about which the resonance angle oscillates differs between the
external and internal perturber cases.  For the pendulum model of the
circular restricted 3-body problem, the equilibrium point is located
at $\phi = 0$ for internal resonances, while for external resonances
it is $\phi = \pi$. We stress, however, that the pendulum model is an
over-simplification for the regime under consideration.

Figure \ref{F:Int_nod} shows the eccentricities (top panel), resonance
angle (middle panel), and phase trajectory (bottom panel) for a system
the same set of initial orbital elements as was used in Figure
\ref{F:Int_anti}, with the exception that here the planet's orbital
eccentricity $e\planet = 0.1$.  The system is prepared with
anti-aligned orbits and the orbiting bodies in conjunction near the
test body's apoapse. Recall from the above discussion that this
particular configuration places the resonance angle near a separatrix
in its phase space. The figure shows slightly more than 3 complete
secular cycles, each lasting $\approx 3300$ yrs, which corresponds to
hundreds of libration times. During each secular cycle, the test
body's orbital eccentricity (shown in black in the top panel)
increases from its initial value of $e = 0.15$ up to $e \sim 0.6$
(where the two orbits intersect for a time), and then decreases close
to its initial value. For times that are shorter than secular
timescales and longer than libration timescales, the resonance angle
undergoes large amplitude oscillation about the test body's periapse.
In the previous case, when the planet's orbit was nearly circular, the
motion of the test body's apsidal angle was steady circulation. In the
present case where the Jovian planet's eccentricity is substantially
larger, the azimuthal symmetry of the star-planet Keplarian system has
been sufficiently broken and the evolution of the test body's
osculating elements depends sensitively upon the orbital alignment
(given by the apsidal angle, $\Delta \varpi$). The red curve in the
middle panel of Figure \ref{F:Int_nod} shows the apsidal angle, which,
over the course of one secular cycle, oscillates once very slowly
about $\Delta \varpi = \pi$, then very abruptly passes $\Delta \varpi
= 0$ in the retrograde direction.  Generally, as the apsidal angle
approaches $\Delta \varpi = \pi$ (i.e., as the orbits approach
anti-alignment), the test body's eccentricity decreases, and as the
orbits rotate out of anti-alignment, the eccentricity increases.  As
the apsidal angle approaches $\Delta \varpi = 0$, the orbits come into
alignment, and with comparatively large orbital eccentricities, the
planet exerts greater influence on the motion of the test body than in
cases where it's orbit is nearly circular (see Batygin \& Morbidelli
2011 for a detailed analysis of secular dynamics).  As a consequence,
the resonance angle circulates once or twice until the orbits rotate
out of alignment, and the system enters into yet another secular cycle
that deviates only slightly from the one just described.  The test
body can exhibit a wide range of eccentricity growth/decay during a
secular cycle, and the duration of the secular cycle depends on the
details of the initial 3-body configuration. However, the generic
behavior of the osculating orbital elements described above during a
secular cycle for any choice of $\Delta \varpi_0$ is robust. As long
as the orbiting bodies are in conjunction with $f_0$ for the test body
near apoapse, the resonance will reside near the separatrix in phase
space and the system can experience nodding for states near mean
motion resonance.

For the external resonance shown in Figure \ref{F:Ext_nod}, the system
is initially configured with anti-aligned orbits $\Delta \varpi = \pi$
and conjunction occurring near the test body's apastron $f_0 \simeq
\pi$. The test body's orbital eccentricity is sufficiently large to
place the system near an asymmetric resonance, meaning there are
generally two libration points.  The exact locations of the libration
points depend on the instantaneous orbital eccentricity (shown by the
black curve in the top panel of the figure), but for ease of
discussion, we take them to be $\approx \pi \pm \pi/2$.  During a
secular cycle, the resonance angle $\phi$ oscillates with an amplitude
$\pi/4 \lesssim \Delta \phi < \pi/2$.  There are 6 full secular cycles
shown here, each lasting $\sim1600$~years. The test body's periapse
circulates in the retrograde direction on average, so the apsidal
angle $\Delta \varpi$ circulates.  The nodding cycle, eccentricity
growth/decay, and the apsidal angle circulation are all governed by
the secular cycle and coordinate together to produce the specific
nodding behavior of the resonance angle.  As the two orbits' major
axes rotate to become more perpendicular, the test body's eccentricity
grows, and resonant oscillation amplitudes decrease, resulting in a
tightening of libration. As the two orbits rotate so that their major
axes become parallel (orbits either aligned or anti-aligned), the test
body's orbital eccentricity decreases.  The eccentricity attains a
minimum value when the major axes are in alignment, reaching its
smallest values when the orbits' periapses become aligned (as opposed
to anti-aligned).  As the test body's orbital eccentricity decreases,
the libration points of the asymmetric resonance move closer to $\pi$
and the resonant angle's oscillation amplitudes generally increase,
which in turn moves the phase trajectory toward one of the two
separatrices. With the orbits near alignment ($\Delta \varpi \approx
0$), the phase trajectory approaches the outer separatrix, and
circulation becomes possible.  When the orbits are nearly
anti-aligned, the phase trajectory approaches the inner separatrix,
which separates bounded libration around a single stable point from
libration around both. Here, it is possible for the phase trajectory
to jump across $\pi$ to the adjacent stable point.  The number of
times the resonant angle circulates or jumps back and forth between
stable points during major axes alignment varies from secular cycle to
cycle. However, the nodding cycles tend to mimic preceding cycles, and
every so often abrupt changes will occur which again persist over
multiple nodding cycles, until another abrupt change occurs. This
behavior then repeats.

To summarize, we have studied both internal and external resonance
scenarios using both nearly circular planet orbits ($e\planet =
10^{-3}$) and an orbit that departs from circular ($e\planet =
0.1$). We took a benchmark value for the test body's initial orbital
eccentricity of $e = 0.15$, but this value varied with the secular
cycle, reaching values as high as $e\simeq 0.6$ and as low as $e\simeq
0.01$.  Variations to the test body's orbital eccentricity increased
in simulations which included larger planet eccentricity.  On secular
timescales, the 2:1 resonance angle may exhibit bouts of circulation
with regular libration in between circulation events (the phenomenon
we call nodding). Nodding is common when the planet's orbital
eccentricity is sufficiently large, where we find an approximate
threshold of $\sim e\planet \gtrsim $~a~few~$\times10^{-2}$. Nodding
systems with larger orbital eccentricities are able to obtain greater
net circulation than systems where both orbital eccentricities are
small. However, the relative orbital alignment and the test body's
true anomaly at the moment of conjunction affect both the period of
the nodding cycle and the libration width during intermediate times. 
Both internal and external resonances exhibit nodding, but prominent
qualitative differences in the nodding signatures between the two
configurations exist.  Because of our choice of initial test body
orbital eccentricity, asymmetric resonances are found for external
cases, where libration occurs around points $\phi = (1 \pm \delta/2)
\pi$ with $\delta \approx 1$ for $e_0 \approx 0.15$ (e.g., Lee 2004;
cf. Callegari et al. 2004). Consequently, for external resonances,
libration amplitudes are typically smaller ($\Delta \phi \lesssim
\pi/4$) for large $e\planet$. For nearly circular planet orbits, large
amplitudes ($\Delta \phi \lesssim \pi$) may persist for external
resonances, and appear to librate about $\phi = \pi$.  However, the
resonance angle's motion slows as it passes $\phi = \pi$, which is
distinctively different from large amplitude oscillations for the
internal resonance case.

\subsection{Dynamical Map of Nodding in the Hot Jupiter Problem}

In this section we explore the nodding phenomenon for Hot Jupiters.
Because of the existence of the asymmetric resonance and the
additional separatrix that comes along with it, the external resonance
is particularly interesting in the context of the Hot Jupiter problem
(where smaller bodies could be found in outer orbits -- see Ketchum et
al. 2011b).  For completeness, we explore how the resonance angles of
the external 2:1 near resonance are affected by the external test
body's orbital eccentricity and by the orbital alignment between the
planet and the test body.  To perform this study, we use typical
parameters of Hot Jupiter systems, where a Jovian planet (with mass
$m\planet = 1M_{jup}$, orbital eccentricity $e\planet = 0.04$, and
orbital period $T\planet = 4$ days) orbits a $1 M_\Sun$ star with a
small planet (with mass $m = 1 M_\Earth$ and orbital period $T = 8$
days) orbiting external to the Star-Jovian system.  We take orbital
eccentricity values for the test body between $-3 \le \log_{10} e \le
-1/3$ in equally spaced logarithmic increments.  For each value of the
test body's eccentricity, we perform an ensemble of 200 similarly
prepared systems, each with a slightly different initial apsidal angle
between $-\pi < \Delta \varpi \le \pi$ in equally spaced
increments. Both the Hot Jupiter and the smaller rocky planet begin
each simulation located at the periastron of their respective orbits
(not necessarily in conjunction with one another). The parameter space
outlined by the above conditions contains 28,560 points, each of which
are integrated for up to 200,000 orbits of the Jovian planet ($\approx
2000$ years). During each individual simulation, we keep track of the
displacement of both the resonance angle $\phi$ and the angle of
apsides $\Delta \varpi$, as well as the total number of times the
resonance angle passes the test body's periastron (the latter provides
a rough measure of the number of circulation events that take place
during the simulation).

\begin{figure}[htbp]
\begin{center}
\includegraphics[width=130mm]{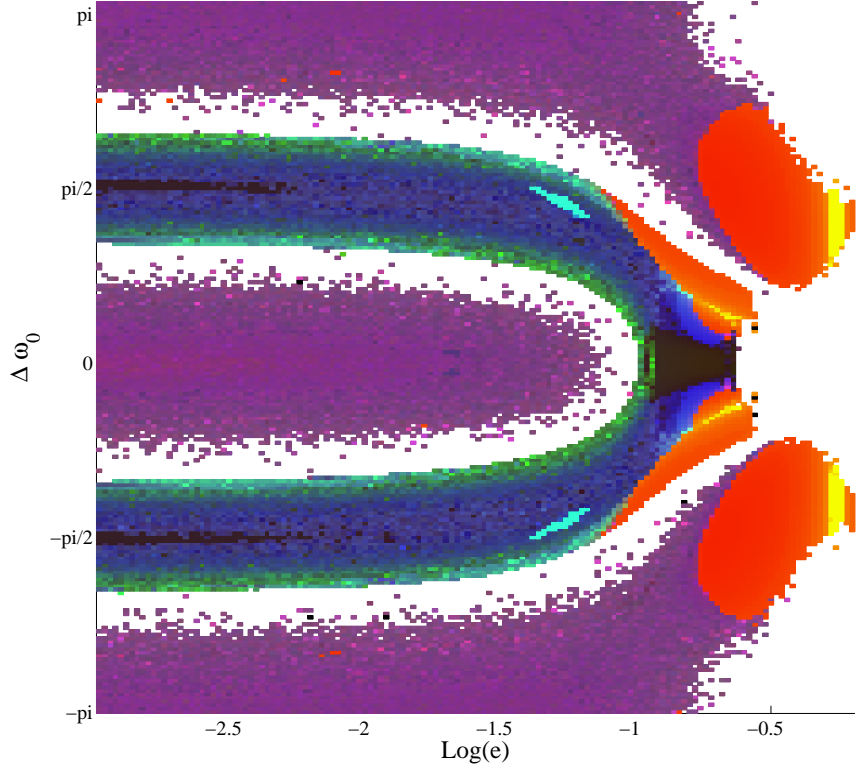}
\caption{Dynamical map of the resonance and apsidal angles' relative  
amounts of accumulated circulation during simulations lasting 200,000
orbits ($\approx 2000$ years) for the external 2:1 near-resonance.
All angles are given in radians.  Each pixel in this figure represents
one simulation with a given initial apsidal angle, $\Delta \varpi$
(vertical axis), and test body orbital eccentricity, $\log_{10} e$
(horizontal axis).  This figure is a mosaic of 28,560 different
simulations performed as described in the text. Each pixel is colored
with an admixture of red, green, and blue. The particular shade of
each color signifies the relative amount of circulation, hence
nodding, using three different metrics -- see text for full
description. This set of simulations was performed under the context
of a Hot Jupiter scenario, i.e., the planet has eccentricity $e\planet
= 0.04$, mass $m\planet = 1$ M$_{jup}$, and period $T\planet = 4$
days, and the test body has mass $m \approx 1 M_\Earth$ and period $T
= 8$ days, both orbiting a $1 M_\Sun$ star.  Each simulation begins
with the planet and test body located at periastron of their
respective orbits.  This figure shows what looks like a pitchfork
bifurcation in this parameter space (perhaps due to the transition
from symmetric to asymmetric resonances following from left to right
across map).  For low test body eccentricities, rapid circulation
occurs for initial apsidal angles $\Delta \varpi = \pm \pi/2$, and as
the initial eccentricity increases, these tracks converge in the
center, where, at $e \approx 0.1$, rapid circulation occurs for states
with an initial apsidal angle $\Delta \varpi \approx 0$.}
\label{F:dyn_map}
\end{center}
\end{figure}

The results of the survey are shown as a dynamical map in Figure
\ref{F:dyn_map}. Each pixel in the figure is composed of an admixture
of the three colors red, green, and blue. The amount of each color
within a given pixel represents different behavioral characteristics
of the angles of interest: \emph{(i)} Red measures the resonances
angle's final angular displacement, \emph{(ii)} Green measures the
angle of apsides final angular displacement, and \emph{(iii)} Blue
measures the resonance angle's final angular displacement in
comparison to the total number of circulation occurrences during the
given simulation.  We use the ensemble averages for each category to
gauge the overall shade of each color of the survey.

Admittedly, our metric for defining each color's particular shade in
each pixel is somewhat arbitrary, so we refrain from providing
specific details about it here. However, in general our metric
produces a bright color for smaller deviations from the initial state
in comparison to the ensemble's average deviation, and a darker shade
corresponds to a larger deviation than average.  We provide a brief
interpretation of the prevalent color combinations presented in the
figure as follows: \emph{RED:} $\phi$ strictly librates (no
circulation events) and $\Delta \varpi$ circulates; \emph{YELLOW:}
Both $\phi$ and $\Delta \varpi$ librate (minimal circulation);
\emph{GREEN:} $\Delta \varpi$ librates, but $\phi$ circulates quickly;
\emph{PURPLE:} $\phi$ circulation events encountered, but direction of
circulation is erratic, and $\Delta \varpi$ ciruclates; \emph{CYAN:}
relatively few circulation events for both $\phi$ and $\Delta \varpi$
and direction of circulation is consistent; \emph{BLACK:} both $\phi$
and $\Delta \varpi$ circulate rapidly; \emph{BLANK:} either \emph{(i)}
premature termination of the simulation due to a scattering or
collision event for the test body, or \emph{(ii)} a period ratio
between the planet and test body that deviates by more than 10\% once
the simulation reaches its time limit.

Although the details appearing in the Figure \ref{F:dyn_map} are
intricate and rich, there two main features that we want to
emphasize here.  [1] The existence of the large, red islands
appearing near $(\log(e), \Delta \varpi_0) = (-0.5, \pi/2)$ and the 
absence of the red islands for  values $\log(e) \lesssim -1$ is in agreement 
with previous findings that for $m_2 \ll m_1$, stable oscillations 
orbits are asymmetric (Beauge 1994). These large red islands are regions of parameter 
space where the resonance angle does not circulate at any point in time 
during the 200,000 orbit simulation. [2] The diagram shows an apparent pitchfork bifurcation
occupying this mapping. For low test body eccentricities, rapid
circulation occurs for initial apsidal angles, indicated by the thin
horizontal tracks of dark (almost black) pixels located around $\Delta
\varpi = \pm \pi/2$.  (For completeness, note that the secular angle
is not defined in the limit $e \to 0$; here the eccentricity axis is
presented on a logarithmic scale so that $e > 0.001$.)  As the
eccentricity is increased, these tracks converge to a single
horizontal line at the center of the map, where at $e \approx 0.1$
rapid circulation of both $\phi$ and $\Delta \varpi$ occurs for
initial apsidal angle $\Delta \varpi \approx 0$. This trend in the
dynamics of the resonance angles, as the exterior test body's
eccentricity is increased, is consistent with past studies (e.g., Lee
2004) and, perhaps, reveals further details about the onset of the
asymmetric exterior 2:1 resonance.  

\newpage

\subsection{Transit Timing Variations in the Presence of Nodding}

One potential channel for discovering terrestrial sized exoplanets
involves observing variations in the transit times for a Hot Jupiter
(e.g., Agol et al. 2005).  These transit timing variations (TTVs) will
be largest if the body responsible for the variations is in (or near)
a mean motion resonance with the Hot Jupiter. Of the multiple-planet
systems known to date, many of the adjacent planet pairs have period
ratios that are close to small integer ratios, with 2:1, 3:2, and 3:1
being the most populated, in order of decreasing frequency (e.g.,
Fabrycky et al. 2012).  These observations suggest that a planet,
whose presence produces transit timing variations of a transiting Hot
Jupiter, is more likely to be near MMR, not deep in MMR with the Hot
Jupiter. On the other hand, planet-pairs could still be deep in
resonance while the ratio of their orbital periods deviates from exact
resonance by as much as $20\%$ (e.g., Batygin \& Morbidelli 2012). As
shown in the previous section, planetary systems near MMR can exhibit
distinctive and interesting behavior --- nodding of the resonance
angle. This phenomenon can lead to interesting signatures in TTVs, 
which are considered in this section. 

As outlined above, in order for nodding to arise, the larger planet
must have nonzero eccentricity. However, Hot Jupiters are subject to
tidal forces, which act to circularize their orbits.  On the other
hand, the observed orbits of Hot Jupiters are often eccentric: The
current sample includes 242 planets with semimajor axes $a<0.1$ AU
(Exoplanets Data Explorer 2012) and 79 of these planets have orbital
eccentricity in the range $0.02<e<0.6$ (and 41 orbits have $e>0.10$).
Some fraction of the Hot Jupiters are thus expected to undergo nodding
(if smaller companion planets are also present). 

The Hot Jupiters could have nonzero eccentricity for several reasons:
[1] The tidal circularization time for 4-day orbits is typically
$\tau_{circ}\sim1$ Gyr (Goldreich \& Soter 1963; Hut 1981), and
observed systems have ages of 2 -- 5 Gyr.  As a result, the initial
eccentricity is decreased by a factor of ``only'' $\sim7-150$. If, for
example, the starting eccentricity $e_0=0.30$, the observed
eccentricity $e\approx0.002-0.04$; younger systems with higher
starting eccentricities can have even larger values.  [2] The tidal
circularization time grows with a high power of semimajor axis,
$\tau_{circ}\sim{a}^{13/2}$ (Hut 1981), so that planets with slightly
larger orbits have much longer circularization times and hence larger
eccentricities. [3] The time required for tidal circularization
depends on the tidal quality factor $Q$ (Goldreich \& Soter 1963),
which depends on the internal structure of the planet.  Since observed
planets are inferred to have a wide range of structures for a given
mass (e.g., Bodenheimer et al. 2003), we expect a wide range of $Q$
values. [4] In multiple planet systems, massive distant planets can
drive Hot Jupiters to have nonzero eccentricities (Adams \& Laughlin
2006).

To demonstrate the possible effects of nodding on transit timing
variations of a Hot Jupiter, we perform additional simulations. We
place two planets in orbit around a $1 M_\Sun$ star -- a $1 M_{jup}$
planet in a 4 day orbit and a $1 M_\Earth$ planet in an 8 day orbit.
The periastra are approximately anti-aligned $\Delta \varpi \approx
\pi$, and the planets are initially placed near conjunction with the
Hot Jupiter at periastron.  The orbits are set to be coplanar. The
initial eccentricity of the smaller planet is taken to be $e$ = 0.15,
whereas the eccentricity of the Hot Jupiter has values $e\planet =
0.1$ and $e\planet = 0.01$. For each choice of $e\planet$, we then
performed numerical integrations using the integration scheme
described above and recorded the times that the center-to-center
displacement between the star and Hot Jupiter were parallel to some
arbitrary line of sight. Neither light travel time nor transit
duration variations are corrected for during the simulations -- a more
rigorous study should take into account the light travel time from the
star to the Hot Jupiter, and should consider the surface-to-surface
grazing chord between the star and Hot Jupiter as the line-of-sight
rather than the center-to-center line (Veras et al. 2011).  However,
these are higher order corrections and are not considered here.

\begin{figure}[htbp]
\begin{center}
\includegraphics[width=160mm]{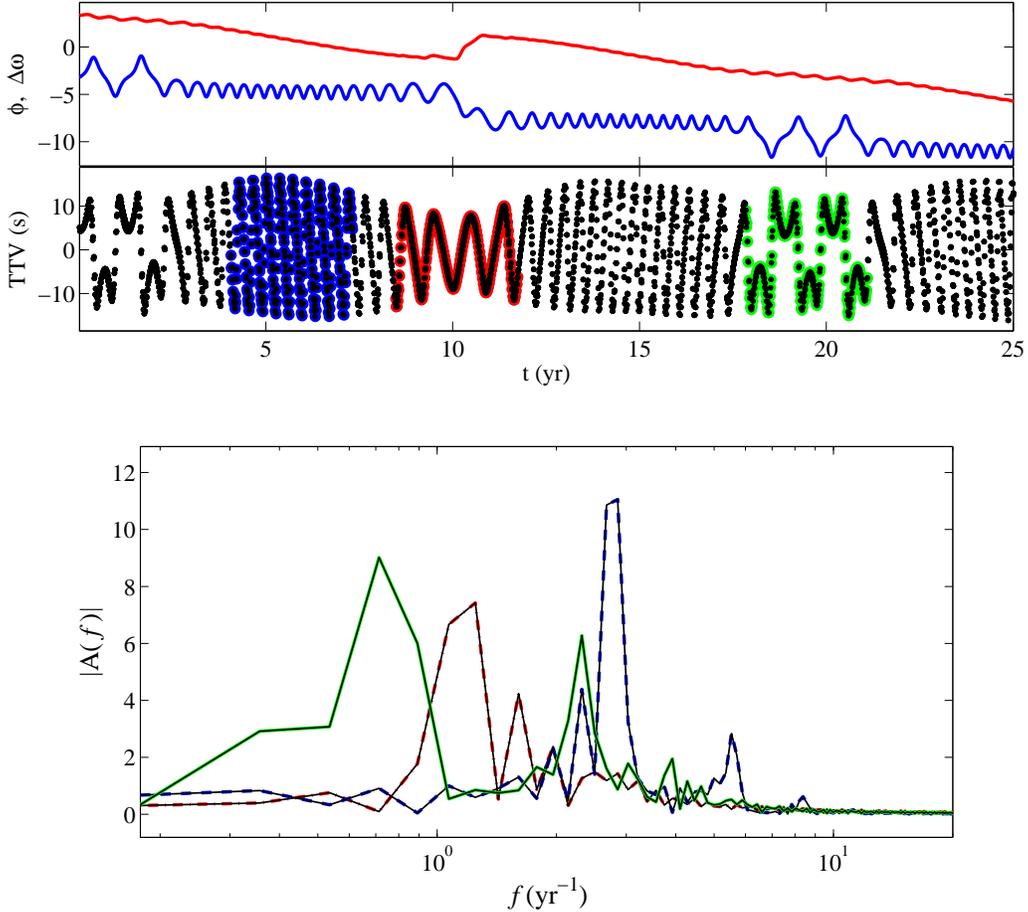}
\caption{Results for a Hot Jupiter in a 4-day orbit, with eccentricity
$e$ = 0.10, and in an external 2:1 near-MMR with an Earth-mass planet
($m \approx 1 M_\Earth$), both orbiting a $1 M_\Sun$ star. The
\emph{top panel} shows the time series for the resonance angle $\phi$
(blue curve) and the angle of apsides $\Delta \varpi$ (red curve) for
the first 25 years ($\approx 2000$ Hot Jupiter orbits). The initial
configuration of this system was chosen to obtain nodding behavior of
the resonance angle. The \emph{middle panel} shows transit timing
variations (in seconds) for the Hot Jupiter during the simulation.  
Three separate 3-year windows highlight the TTV data (corresponding to
different nodding modes of the resonance angle) in the colors (from
left to right) blue, red, and green. Fourier Transforms for each
highlighted portion are shown in the \emph{bottom panel}, and are
color coded to match their corresponding progenitive TTV signal. The
bottom panel demonstrates that the fundamental frequency of the TTV
signal for a system undergoing nodding motion can vary by a factor
$\sim 2$ (either way) depending on resonance angle nodding effects.}
\label{F:ttv}
\end{center}
\end{figure}

Figure \ref{F:ttv} shows the results of one such simulation for the
larger eccentricity $e\planet$ = 0.10 of the Hot Jupiter.  The
resonance angle, shown in blue on the top panel, begins in one mode of
oscillation, is caught in libration while being passed back and forth
between fixed points near $-\pi/2$ and $-3 \pi/2$, and then settles on
the $-3\pi/2$ fixed point for nearly 20 librations.  At $t = 10$ years
($\approx 1000$ orbits), the periastra have rotated into near
alignment $\Delta \varpi \approx 0$, and the resonance angle undergoes
a different type of nodding than exhibited at the beginning, a mode
which is characterized by free circulation across the unstable point
located at $\phi = -2\pi$. Near the $t$ = 12 year mark, the resonance
angle enters into a mode of libration around a single fixed libration
point at $\phi \simeq -5\pi/2$.  This libration undergoes around 20
oscillations, at which time the cycle starts over from the beginning
and is repeated several times. The particular cycle illustrated here,
which produces the nodding pattern, is similar to that exhibited by
the system represented in Figure \ref{F:Ext_nod}; the initial orbital
configurations are nearly identical. However, the mass of the smaller
body for this case, with $m = 1 M_{\Earth}$, is much larger and the
two planets are located much deeper inside the stellar potential than
in the previous case, deep enough that stellar damping plays a role in
the dynamics. These differences contribute to producing the different
signatures of the nodding patterns for the two cases.

The transit timing variations described above and shown in Figure
\ref{F:ttv} exhibit many different modes of oscillation. In addition,
this figure shows that it is plausible that the transit timing
variations closely follow the time derivative of the resonance angle.
When the resonance angle is nodding between the fixed points on either
side of $\pi$, the derivative $\dot{\phi}$ decreases as $\phi$ passes
$\pi$, and then increases before $\phi$ reaches the stable point
opposite $\pi$ and reverses direction.  This type of motion
qualitatively matches the transit timing variations calculated during
the same period of time. When the resonance angle increases, the
transit timing variations are positive; when the angle $\phi$
decreases, the TTVs are negative; and when the angle passes through
$\pi$, the angle slows down and the TTVs decrease. This complicated
behavior is depicted by the 'double-peaked' patterns in Figure
\ref{F:ttv} between $t=0-2$ years and $t=20-25$ years. 

The bottom panel in Figure \ref{F:ttv} shows the Fourier Transforms
(FT) of the TTV signal in 3 year windows which encapsulate times
corresponding to the three different modes of libration exhibited by
the resonance angle's nodding cycle.  The three different power
spectra in the bottom panel are color coded and correspond to the
regions in the TTV signal highlighted with the same color.
Accordingly, the dashed blue curve is the FT for the portion of the
TTV signal in the middle panel that is highlighted in blue, near $t
\sim 5-8$ years.  During this window, the resonance angle librates
around a single fixed point, and the peak frequency given by the FT is
$f \sim 3$ yr$^{-1}$.  The red FT curve is from the TTV signal in the
time window of $t \sim 10-13$ years, which corresponds to a dynamical
state of the system which exhibits a net resonance angle circulation
with aligned orbits.  The peak frequency of the red FT curve is
$f\sim1.4$ yr$^{-1}$. Finally, the green FT curve from the TTV signal
between $t \sim 21-24$ years has a broad peak near frequencies
$f \sim 0.5-0.8$ yr$^{-1}$.  Thus, for this particular system,
observations of TTVs over a three year time interval can yield a range
of fundamental frequencies that span a factor of $\sim 4$; the result
depends on the three year window for which the transit times are
observed.

\begin{figure}[htbp]
\begin{center}
\includegraphics[width=160mm]{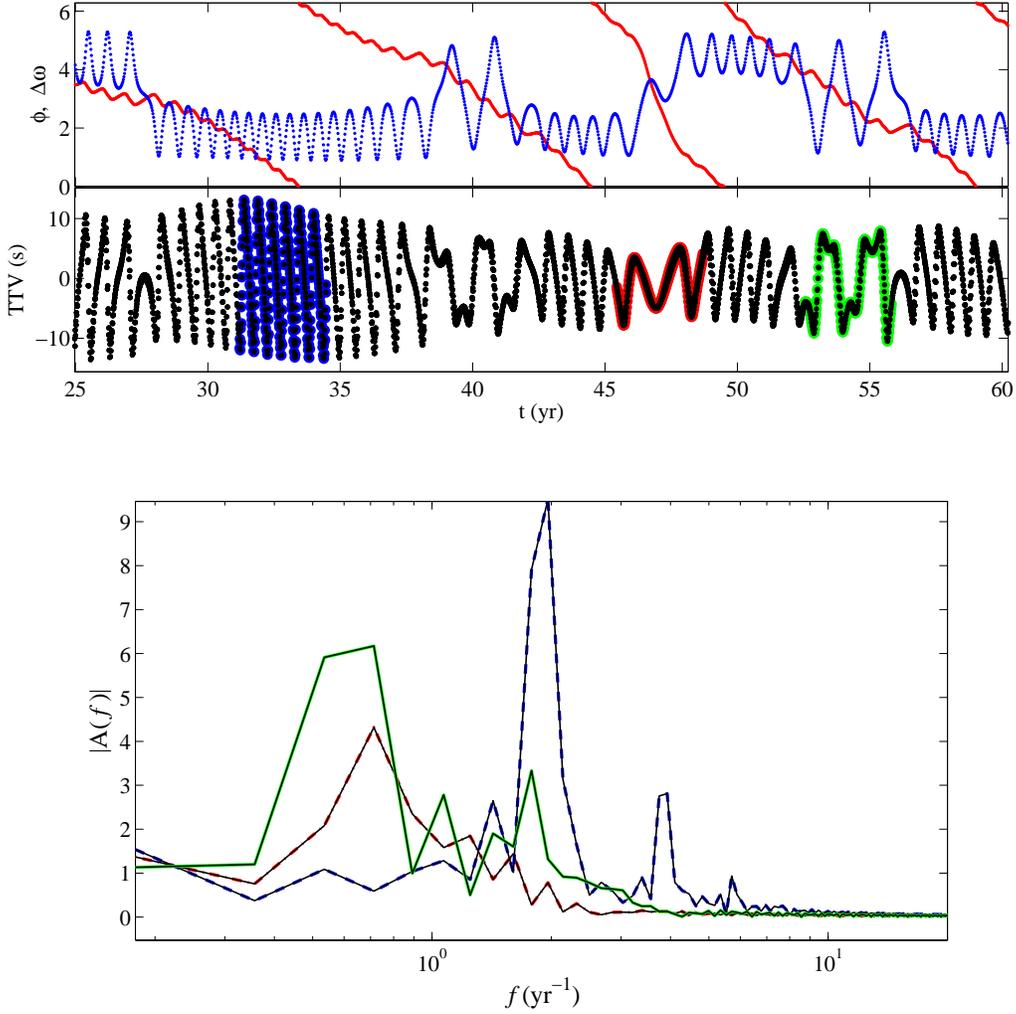}
\caption{Results for a Hot Jupiter in a 4-day orbit, with eccentricity
$e$ = 0.01, and in an external 2:1 near-MMR with an Earth-mass planet
($m \approx 1 M_\Earth$), both orbiting a $1 M_\Sun$ star. The
\emph{top panel} shows the time series for the resonance angle $\phi$
(blue curve) and the angle of apsides $\Delta \varpi$ (red curve) for
$\approx 35$ years ($\approx 3000$ Hot Jupiter orbits). The initial
configuration of this system was chosen to obtain nodding behavior of
the resonance angle. The \emph{middle panel} shows transit timing
variations (in seconds) for the Hot Jupiter during the simulation.  
Three separate 3-year windows highlight the TTV data (corresponding to
different nodding modes of the resonance angle) in the colors (from
left to right) blue, red, and green. Fourier Transforms for each
highlighted portion are shown in the \emph{bottom panel}, and are
color coded to match their corresponding progenitive TTV signal. The
bottom panel demonstrates that the fundamental frequency of the TTV
signal for a system undergoing nodding motion can vary by a factor
$\sim 2$ depending on resonance angle nodding effects.}
\label{F:ttv2}
\end{center}
\end{figure}

Figure \ref{F:ttv2} shows the time evolution for a planetary system
that is identical to that shown in Figure \ref{F:ttv}, except that the
Hot Jupiter is taken to have smaller eccentricity $e\planet=0.01$.
The behavior of the resonance angle is much the same as before, i.e.,
the transit timing variations calculated for the Hot Jupiter in this
case contain all of the same qualitative features discussed in the
previous example. For completeness, we have run additional scenarios
where the mass of the smaller planet takes on different values. The
resulting time evolution of the resonance angles (not shown) are,
again, qualitatively similar. As the mass of the smaller body varies,
however, the amplitude of the transit timing variations change in
proportion, as outlined below (see also Agol et al. 2005; Veras et
al. 2011).

Before leaving this section, we derive an approximate relation between
TTV amplitudes (for the Hot Jupiter scenarios outlined above) and the
time rate of change for the resonance angle $\dot{\phi}$.  The
applicable resonance angle for this situation is given in equation
(\ref{E:phi_ext}).  The second time derivative for the resonance angle
is given by
\be\label{E:dotphideriv}
\ba
\ddot{\phi} &= (p+q)(\dot{n}+\ddot{\ueps})_\Earth - 
p(\dot{n} +\ddot{\ueps})_{j} - q \ddot{\varpi}_\Earth \\
		&\approx  (p+q)\dot{n}_\Earth - p \dot{n}_{j} \ .
\ea
\ee 
Lagrange's planetary equations of motion provide us with the time rate
of change for the mean motion for both the Hot Jupiter and the Super
Earth in this problem,
\be \label{E:nsedot}
\dot{n}_\Earth = -\frac{3}{a_\Earth^2}(p+q) \Dp \R_I \ ,
\ee
\be\label{E:nhjdot}
\dot{n}_{j} = \frac{3}{a_{j}^2}p \Dp \R_E \ ,
\ee
where $\R_I$ and $\R_E$ are the disturbing functions for an internal
and external perturber, respectively.  Using a simple, time averaged
form for the internal disturbing function,
\be
\R_I = \frac{\mathcal{G} m_{j}}{a_{j}} 
\left[ \alpha \mathcal{S} + \FI e_\Earth \cos{\phi} \right] \ ,
\ee
and the external disturbing function,
\be
\R_E = \frac{\mathcal{G}m_\Earth}{a_\Earth} 
\left[ \mathcal{S} + \FE e_\Earth \cos{\phi} \right] \ ,
\ee
where $\mathcal{S}$ is the secular contribution to the disturbing
function, we substitute these forms into equations (\ref{E:nsedot})
and (\ref{E:nhjdot}) and get a relation between $\dot{n}_\Earth$ and
$\dot{n}_{j}$,
\be \label{E:nrelation}
\dot{n}_j = -\mu \frac{p}{p+q}\alpha^3 \frac{\FE}{\FI} \dot{n}_\Earth \ ,
\ee
where $\mu \equiv m_\Earth/ m_j$ is the ratio of planet masses. In the
case of a $2:1$ resonance, $\alpha \approx 2^{-2/3}$, and the portion
of equation (\ref{E:nrelation}) that depends on $\alpha$ simplifies to
\be
\alpha^3 \frac{\FE}{\FI} \simeq \alpha^3 
\frac{f_{31} - 2 \alpha}{\alpha f_{31} - (2\alpha)^{-1}} = \alpha^2 \ .
\ee
Substituting this simplification into equation (\ref{E:nrelation}) and
combining the result with the approximate form for the second time
derivative in equation (\ref{E:dotphideriv}), we find an expression
relating the time rate of change for the Hot Jupiter's orbital period
to the second time derivative of the resonance angle,
\be\label{E:phiddotintegrate}
\frac{\dot{T}}{T^2} \approx \frac{\mu}{2^3\pi \alpha^2} \ddot{\phi}\ ,
\ee
where we have made use of the relation $\dot{n} = -2 \pi T^2 \dot{T}$
in this last line.  Neglecting time variations in $\alpha$, equation
(\ref{E:phiddotintegrate}) can be directly integrated (over one orbit
of the Hot Jupiter) to give us the desired relation between transit
timing variations, $\Delta T$, and the time derivative of the
resonance angle, $\dot{\phi}$,
\be\label{E:ttv}
\Delta T \approx \frac{\mu T_j^2}{2^{5/3}\pi} \dot{\phi} \ .
\ee

To demonstrate the accuracy of equation (\ref{E:ttv}), we take Figure
\ref{F:ttv} as an example.  At peak TTV in Figure \ref{F:ttv} (near
the 5 year mark in the TTV panel of the figure), the resonance angle
is changing at a rate of $\approx 12$ rad yr$^{-1} \sim 3.8 \times
10^{-7}$ rad sec$^{-1}$.  The Hot Jupiter's period is roughly 4 days
($\sim 3.5 \times 10^5$ sec), and the outer planet has mass $m$ = 1
$M_\Earth$ ($\mu \approx 0.003$).  Substituting these quantities into
the right hand side of equation (\ref{E:ttv}) gives a timing variation
$\Delta T \approx 14$ sec, a result which is in good agreement with
the figure. This result shows that the TTV signal can reveal intricate
details about near-resonance nodding if such a system can be found in
nature.  The above derivation also confirms that the strength of the
TTV signal increases with the mass of the exterior planet (Agol et al.
2005). We have run additional simulations with varying masses for the
outer planet, up to $m$ = 10 $M_\Earth$, and find at the amplitudes of
the TTVs vary in proportion to the mass (as expected). In addition,
equation (\ref{E:ttv}) shows that the TTV signal increases as the
square of the Hot Jupiter's orbital period and one power of
$\dot{\phi}$. Since $\dot{\phi}$ itself is inversely proportional to
the period, the TTV amplitudes are directly proportional to the
orbital period.

\newpage
\section{Derivation of Model Equations for Nodding} \label{S:eom}

In this section, we turn our attention to a Lagrangian formalism for
the scenarios outlined in section 2. We analyze the equation of motion
for the resonance angle in the restricted 3-body problem in order to
identify the terms that encapsulate the dynamics of nodding. We choose
a Lagrangian formulation instead of a Hamiltonian treatment purely for
ease of physical interpretation. Consider three point masses that
follow a hierarchical arrangement, $m\testbod \ll m\planet \ll
m\cntr$, where we denote orbital elements belonging to the test body
with subscript $\TESTBOD$ and those belonging to the planet with
$\PLANET$. The three bodies are spatially separated so that
$m\testbod$ and $m\planet$ are gravitationally bound to $m\cntr$, but
not bound to one another. Furthermore, suppose the orbits for
$m\testbod$ and $m\planet$ are coplanar, so they orbit $m\cntr$ in the
same plane. In what follows, first we provide a general treatment of
the equations where we do not specify whether the planet's orbit is
interior or exterior to the test body's orbit, but will later focus on
the internal perturber case where $a\planet < a\testbod$. The system
sketched in the above outline is the planar restricted 3-body
problem. This coming derivation follows a mathematical approach
similar to that used in the text by Murray \& Dermott (1999).

The argument angles (and hence the resonance angle), $\phi$, for the
disturbing function in the planar 3-body problem are given by a linear
combination of the Eulerian angles
\be\label{E:phi}
\phi = j\testbod\lambda\testbod + j\planet\lambda\planet + k\testbod\varpi\testbod +k\planet\varpi\planet \, ,
\ee
where $j\testbod$, $j\planet$, $k\testbod$, and $k\planet$ are
integers, which sum to 0, $j\testbod + j\planet + k\testbod + k\planet
= 0$. For such a system in resonance, $|j\planet|:|j\testbod|$ is a
ratio of small integers, and $j\planet$ and $j\testbod$ are of
opposite sign.  From simulations, we found that the angle that tends
to librate first is the angle corresponding to that of the circular
restricted 3-body problem, i.e., where $k\planet \equiv 0$ in the case
where the perturbing planet's orbit is strictly circular.  This is
because the strength functions associated with the resonance angle
(besides those of the secular angles) are proportional to $k\planet$
factors of the planet's orbital eccentricity $e\planet$ and
$k\testbod$ factors of the test body's orbital eccentricity
$e\testbod$. The perturbing planet and the primary mass (star) form a
nearly Keplarian system. As a result, we assume that
$\dot{\lambda}\planet \approx n\planet$ in this derivation. From this
point forward, we note that orbital elements appearing without a
subscript belong to the test body.

We find the first time derivative for the resonance angle 
\be\label{E:Phidot}
\dot\phi = j\testbod \left(n + \underline{\dot\epsilon} \right) + j\planet n\planet - (j\planet + j\testbod) \dot\varpi\ ,
\ee
where the quantity
\be\label{E:ueps}
\dot{\ueps} = \dot\epsilon + \dot{n}\, t \
\ee
is introduced to avoid explicit time occurrences in the equation of motion (Brouwer \& Clemence 1961) and where $\dot{n}$ in equation (\ref{E:ueps}) may originate from both damping effects and the disturbing function.  In the above derivation, it is important to note a few things. From this point forward we use the following notation; partial derivatives written as $\partial_q$ operate only on explicit occurrences of the variable $q$, and otherwise vanish.  
Using a lowest eccentricity order form of Lagrange's planetary equations of motion for $\dot{\ueps}$ and $\dot{\varpi}$, we can write
\be\label{E:pomdot}
\dot{\varpi} = \frac{1}{na^2e}\De\mathcal{R} \ ,
\ee
and
\be\label{E:epsdot}
\dot{\ueps} = -\frac{2}{na}\partial_a \mathcal{R} + \frac{e}{2na^2}\De \mathcal{R} =
\sA\frac{2 \alpha}{n a^2} \Da \mathcal{R} + \frac{e^2}{2} \dot{\varpi} \ ,
\ee
where $\mathcal{R}$ is the disturbing function (for either an interior or an exterior perturber), and where we introduce $\sA$ with $\sA = \{ 1, -1\}$ for an internal or external perturber, respectively; this sign function arises from a change of variables from $a\testbod$ to $\alpha$, i.e., $a\partial_a = -\sA \alpha \Da$. We can rewrite equation (\ref{E:Phidot}) as
\be\label{E:PHIDOT}
\dot{\phi} = \left(j\testbod + \mu \alpha^{\sB} \h (\R) \right)n +j\planet n\planet \ ,
\ee
where $\mu = m\planet / m\cntr $ is the mass ratio between the planet and the star, we introduce the quantity $\sB$ where $\sB = \{ 0, 2 \}$ for an internal or external perturber, respectively; we have also introduced the linear differential operator $\h$ defined as
\be\label{E:H}
\h \equiv  \Lambda^{-1} (\ah\Da - \eh \De) \ .
\ee
In equation (\ref{E:H}), $\Lambda \equiv \G m\planet / a\planet$ is the pre-factor to the disturbing function containing dimensionful constants, and we introduce the quantities $\ah$, $\eh$, and $\beta$ defined by 
\be\label{E:H_aux}
\ba
\ah &\equiv 2 j\testbod \ \sA \ , \\
\eh & \equiv \frac{(j\testbod + j\planet) \beta}{\alpha e} \ , \\
\text{and} \qquad \beta & \equiv 1 - \frac{j\testbod}{j\testbod+j\planet}\frac{e^2}{2} \ .
\ea
\ee
Note that when substituting Lagrange's planetary equation of motion for $\dot\ueps$ in equation~(\ref{E:epsdot}), we have kept contributions from variations to the disturbing function from $\alpha$, whereas some treatments will ignore these contributions (e.g., MD99), equivalently asserting that $\Da~\simeq~0$. However, this contribution can be substantially larger (nearly 40 times greater) than the contribution to $\dot{\ueps}$ from the $\dot{\varpi} e^2/2$ term. The form in equation (\ref{E:PHIDOT}) makes it easier to take the second time derivative en route to the resonance angle's equation of motion,
\be\label{E:ddotphi_quick}
\ddot{\phi} = \left[j\testbod+ \mu \alpha^{\sB} \h (\R)\right] \dot{n} + \mu n \alpha^{\sB}\left[ \sB \frac{\dot{\alpha}}{\alpha} \h(\R) + \frac{d}{dt}\left\{ \h(\R) \right\} \right]\ .
\ee
Using the form of the operator $\h$ given in equation (\ref{E:H}), we can apply the chain rule for the time derivative in the last term of equation (\ref{E:ddotphi_quick}) to obtain
\be
\frac{d}{dt}\left[ \h(\R) \right] = \dot{\h} (\R) + \h (\dot{\R})  \ .
\ee
The time derivative of the $\h$ operator, for constant primary and secondary masses, and constant semi-major axis for the secondary mass, is given by
\be
\dot{\h}(\R) ={\Lambda}^{-1} \eh  \left[ \frac{\beta^{*}}{\beta}  \left(\frac{\dot{e}}{e}\right) + \sA \frac{2}{3} \left(\frac{\dot{n}}{n}\right) \right] \De\R \ ,\quad \text{with} \quad \beta^{*} \equiv 2-\beta  \ ,
\ee
and 
\be
\h(\dot{\R}) = \dot{e} \h(\De \R) + \sA \frac{2}{3} \frac{\dot{n}}{n}\alpha \h(\Da \R) + \dot{\varpi} \h(\Do \R) + \dot{\phi} \h(\Dp \R)\ .
\ee
Combining the previous two results with equation (\ref{E:ddotphi_quick}) gives
\be\label{E:phiddot_d}
\ba
\ddot{\phi} &= \frac{\dot{n}}{n} \left[
			\dot\phi - j\planet n\planet
		     + \sA \frac{2}{3} \mu n \alpha^{\sB} \left\{ \sB \h (\R) + \alpha \h (\Da \R) +\Lambda^{-1}  \eh \De \R \right\}
			\right]\\
		& \quad
		+\mu n \alpha^{\sB} \left[\frac{\dot{e}}{e} 
		  \left(
		 e \h(\De \R) + \Lambda^{-1} \eh \frac{\beta^*}{\beta} \ \De \R
		 \right)
		  +\dot{\phi} \h (\Dp \R)
		 +\dot{\varpi}  \h (\Do \R)\right]\ .
\ea
\ee
Equation (\ref{E:phiddot_d}) is arranged so that disk damping effects can be easily included into the equation of motion -- all time derivative operations of orbital elements have been collected and accounted for and occur as first order time derivatives. With this form, one can simply substitute Lagrange's planetary equations of motion where necessary and remove all implicit instances of time from the equation of motion, i.e., all remaining operators act only as partial derivatives with respect to the orbital elements, independent of time.
Along these lines, one could modify the time derivatives of the orbital elements to include external forces (e.g., damping)
\be \label{E:damping}
\ba
\dot{n} &= (\dot{n})\dist + (\dot{n})\damp\, , \\
\dot{e} &= (\dot{e})\dist + (\dot{e})\damp\, ,
\ea
\ee
where $(\dot{q})\dist$ represent variations to the orbital element $q$ due to the disturbing function, and $(\dot{q})\damp$ are variations caused by external forces due to interactions with, for instance, a circumstellar disk.

\subsection{Specialization to the Case of External resonances}

We will now focus an external  $p+q: p$ resonance. For a given angle $\phi$ from equation (\ref{E:phi}), The simplest time averaged disturbing function (to second order in eccentricity) for an internal perturber takes the form
\be\label{E:dist}
\mathcal{R}_{\phi} = \Lambda \left[ \alpha \left( \mathcal{S}_{1} + \mathcal{S}_{2} \cos{(\varpi -\varpi\planet)}\right)+  \mathcal{R}_D \cos{\phi} \right]  \, ,
\ee
where
\be\label{E:shorthand}
\ba
\Lambda &\equiv \frac{\mathcal{G} m\planet}{a\planet} \, ,\\
\mathcal{S}_{1} &\equiv  \left(e^2 + e\planet^2\right)f_{s,1}(\alpha) \, ,\\
\mathcal{S}_{2} &\equiv  ee\planet f_{s,2}(\alpha) \, , \\
\mathcal{R}_D &\equiv e^q \FI(\alpha)  \quad \mathrm{where} \quad \FI(\alpha) \equiv (\alpha f_d(\alpha) + f_i(\alpha))\ ,
\ea
\ee
(MD99).
Under this scenario, the following quantities will have values $j\testbod = (p+q)$, $j\planet = -p$,  $\sA = 1$, and $\sB = 0$. Using the damping forms given above in equation (\ref{E:damping}), we rearrange equation (\ref{E:phiddot_d}) algebraically and use the time averaged form of the disturbing function in equation (\ref{E:dist}), and parse the equation of motion into generalized reoccurring operators, trigonometric functions of $\phi$ and $\Delta \varpi$, and damping terms. Note that, during this expansion, we expand the equation for $\dot{e}$ to the second lowest order in eccentricity to retrieve a form that utilizes the parameter $\beta$.  This procedure results in an equation of motion of the form
\be\label{E:eomcompact}
\ba
\ddot{\phi}  = -&\sin \phi \left[ \N pn\planet + (\N  + \mu n \Lambda \h(\Rd))\dot{\phi} + \J(\alpha \s) + \J (\alpha \sS) \cos\pomdiff + \J(\Rd) \cos\phi\right] \\
  -&\sin\pomdiff \left[ \mu n \dot\varpi \Lambda \h(\alpha \sS) + \kk(\alpha \s) + \kk(\alpha \sS) \cos\pomdiff + \kk(\Rd) \cos\phi\right] \\
 +& \left[ \tilde\N (pn\planet + \dot{\phi})  + \LL(\alpha \s) + \LL(\alpha \sS) \cos \pomdiff + \LL(\Rd) \cos \phi\right] \ .
\ea
\ee
Here, new operators were introduced to compactify notation
\be
\ba\label{E:eom_operators}
\J(\Q)&\equiv \N \hn(\Q) + \Ep \he(\Q)\ ,\\
\kk(\Q) &\equiv \Ew\he(\Q)\ , \\
\LL(\Q) &\equiv \tN \hn(\Q) + \tE\he(\Q) \ ,
\ea
\ee
with
\be \label{E:intermediate_operators}
\ba
\hn (\Q)& \equiv \frac{2}{3} \mu n \left[ \alpha \Lambda \h(\Da \Q) + \eh \De \Q \right] \ , \\
\he (\Q) & \equiv \mu n\left[ e \Lambda \h(\De \Q) + \eh \frac{\beta^*}{\beta} \De \Q \right] \ ,
\ea
\ee
where $\Q$ is some function of eccentricities and $\alpha$ given in equation (\ref{E:shorthand}),
\be
\ba \label{E:auxcoeff}
\N &\equiv  -3\ \C (p+q)\Rd \ , \\
\Ew & \equiv  -\C\frac{\alpha}{e^2}\sS \ ,\\ 
\Ep &\equiv  \ \C\frac{q\beta}{e^2} \Rd \ ,
\ea
\ee
where $\C = \mu n / \alpha$, and finally 
\[
\tN \equiv  \left(\frac{\dot{n}}{n}\right)\damp \quad \text{, and} \quad
\tE \equiv  \left(\frac{\dot{e}}{e}\right)\damp \ . 
\]
Each line of equation (\ref{E:eomcompact}) can be analyzed to determine what contributions to the resonance angle's equation of motion come from the disturbing function's secular and direct portions ($\s$, $\sS$, and $\Rd$ from equation (\ref{E:dist})) and what contributions might be expected to arise from external damping forces. The first line of equation (\ref{E:eomcompact}) contains the immediate contributions from the direct part of the disturbing function -- that is to say that each term inside the square brackets includes a factor of $\Rd$, but also included are higher order cross term contributions to the motion arising from secular effects on $\Rd$.  The second line contains the contribution from the secular parts of the disturbing function -- for the \emph{circular} restricted three body problem, this line vanishes. The final line of equation (\ref{E:eomcompact}) originates from the presence of damping forces like those introduced in equation (\ref{E:damping}).

\begin{figure}[htbp]
\begin{center}
\includegraphics[width=150mm]{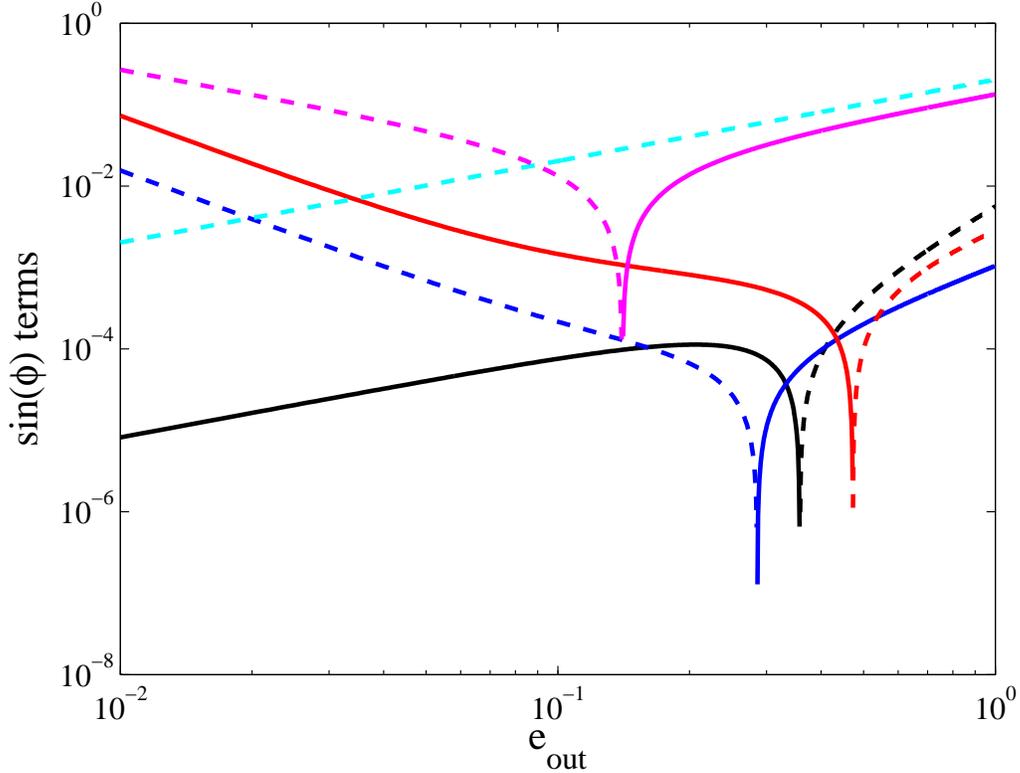}
\caption{Magnitudes of coefficients for terms appearing in the first line ($\sin \phi$ part) of equation (\ref{E:eomcompact})  as a function of test body eccentricity $e$. Solid curves represent positive values, and dashed curves negative.  The magenta curve is the coefficient to $\dot\phi$ (second term in square brackets of line 1 of the equation).  The cyan curve is the usual pendulum term (first term). The red curve is $\J(\Rd)$, the coefficient to $\cos \phi$ (last term). The blue curve is $\J(\alpha \sS)$, the coefficient to $\cos \pomdiff$ (fourth term). The black curve is $\J (\alpha \s)$ (third term).  The figure shows that, for low values of eccentricity, the pendulum term is not necessarily the dominant term. The figure depicts values for 2:1 external resonances, with $n = 2 \pi$, $\mu = 10^{-3}$, and $e\planet = 0.1$.}
\label{F:sinphi}
\end{center}
\end{figure}

\begin{figure}[htbp]
\begin{center}
\includegraphics[width=150mm]{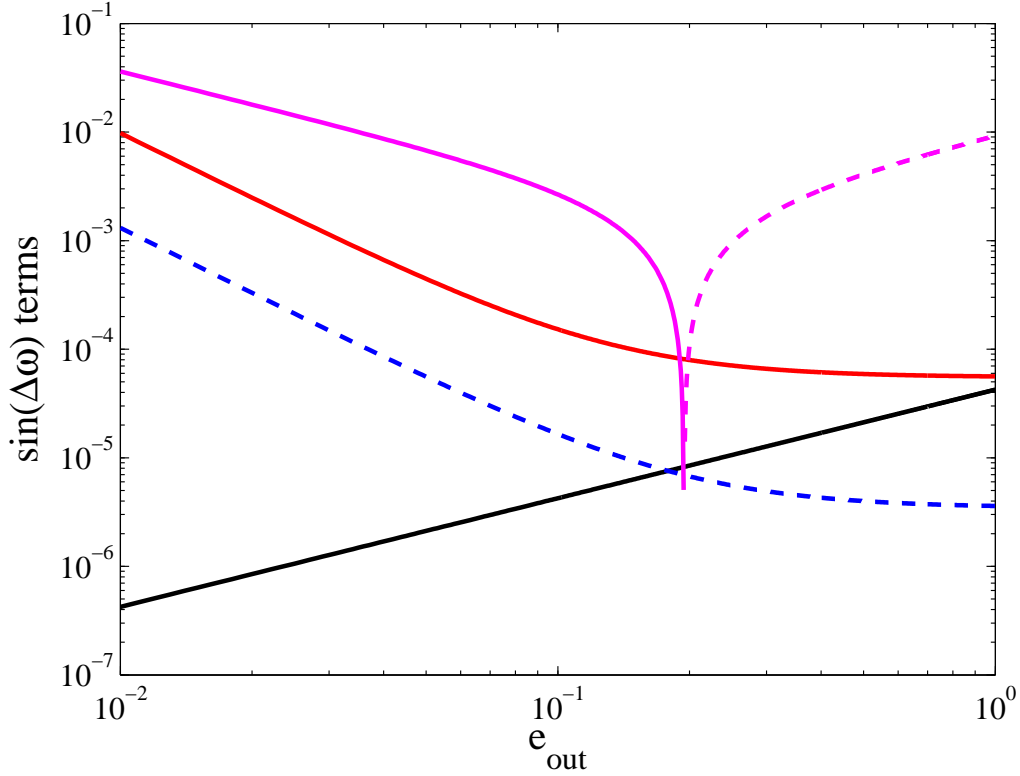}
\caption{Magnitudes of coefficients for terms appearing in the second line ($\sin \Delta \varpi$ part) of equation (\ref{E:eomcompact}) as a function of test body eccentricity $e$. Solid curves represents positive values, and dashed curves negative. The magenta curve is the coefficient to $\dot{\varpi}$ (first term in square brackets of line 2 of the equation). The red curve is $\kk(\Rd)$, the coefficient to $\cos \phi$ (last term). The blue curve is $\kk(\alpha \sS)$, the coefficient to $\cos \pomdiff$ (third term). The black curve is $\kk(\alpha \s)$ (second term).  The only term independent of additional contributions from the angles ($\phi$ and $\Delta \varpi$) in this figure is the black curve This figure depicts values for 2:1 external resonances, with $n = 2 \pi$, $\mu = 10^{-3}$, and $e\planet = 0.1$.  The relative magnitudes are dependent on different factors of $e\planet$.  All strengths vanish for $e\planet = 0$, i.e., for the case of the circular restricted 3-body problem.}
\label{F:sindpom}
\end{center}
\end{figure}

\begin{figure}[htbp]
\begin{center}
\includegraphics[width=150mm]{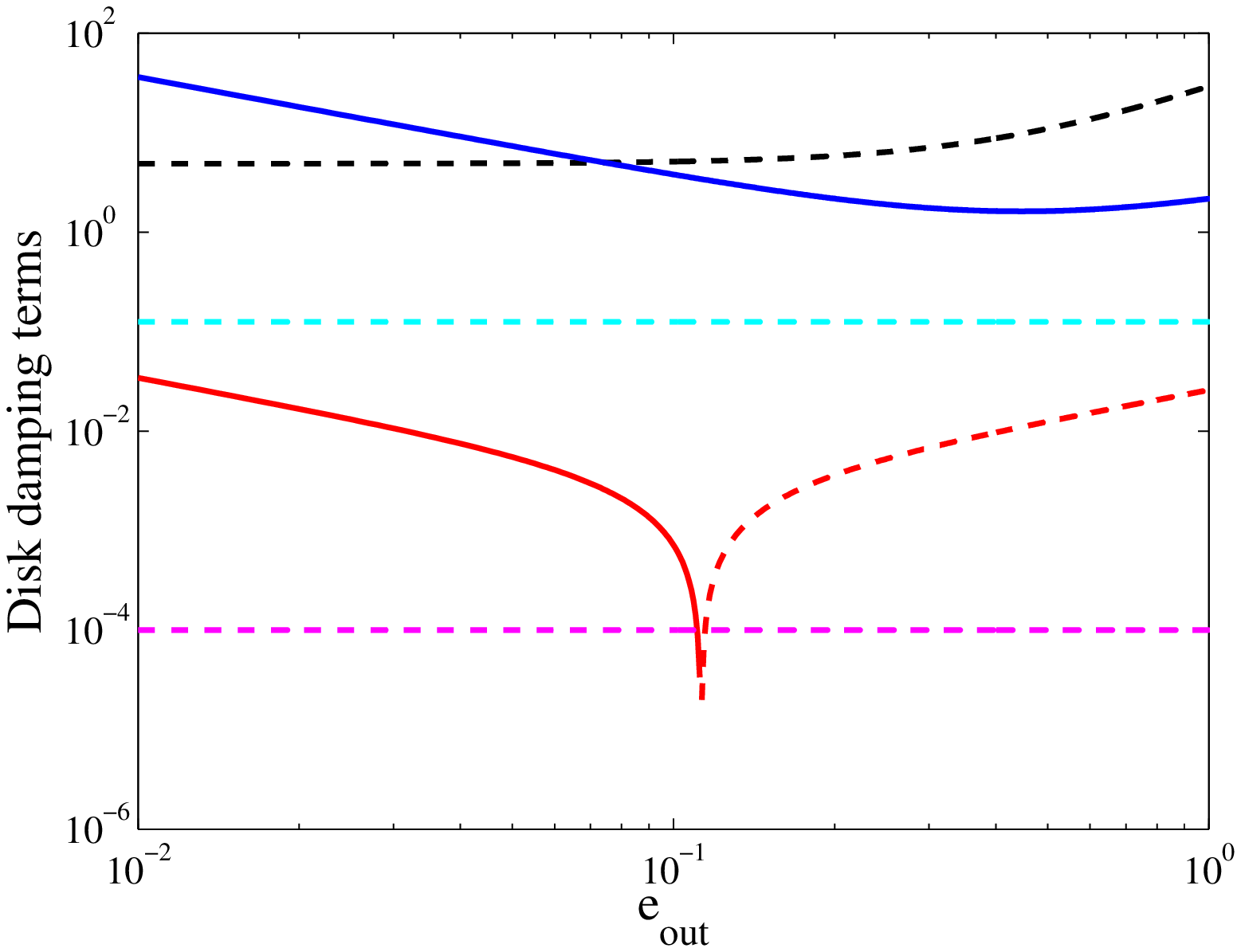}
\caption{Magnitudes of coefficients for terms appearing in the damping terms (third line of the equation) of equation (\ref{E:eomcompact}) as a function of test body eccentricity $e$, e.g., damping terms that may be considered as contributions to $\dot{n}$ and $\dot{e}$ (in addition to the usual Lagrange planetary equations) as in equation (\ref{E:damping}). Solid curves represent positive values, and dashed curves negative. The magenta curve is the coefficient to $\dot{\phi}$ (the first term). The red curve is $\LL(\Rd)$, the coefficient to $\cos \phi$ (fourth term). The blue curve is $\LL(\alpha \sS)$, the coefficient to $\cos \pomdiff$ (third term) -- it is not present in the circular restricted 3-body problem. The black curve is $\LL(\alpha \s)$ (second term). The cyan curve is $\tN p n\planet$. This figure depicts values for 2:1 external resonances, with $n = 2 \pi$, $\mu = 10^{-3}$, and $e\planet = 0.1$. Damping parameters are $\tau_a = 10^4$ years and $\tau_e= 10^3$ years, where we have assumed a simple definition for damping parameters $\dot{a}/a = -\tau_a^{-1}$ and  $\dot{e}/e = - \tau_e^{-1}$. }
\label{F:damp}
\end{center}
\end{figure}

We can expand each operator and determine to what order the eccentricity contributes to the overall magnitude in each individual term therein:  
\be\label{E:JaQ}
\J(\alpha \Q) = - \C^2 \Rd
	\left[ 
		\ah^2\Da(\alpha^2\Da\Q)
		-
		2\alpha^2 \ah\eh\De\Da\Q
		+
		\alpha^2 \eh^2\De^2\Q
		-
		\left(
			\frac{q\beta^{*}}{e^2}
			+
			\ah
		\right)
		\alpha \eh\De\Q
	\right]\ ,
\ee
\be\label{E:JQ}
\J(\Q) = -\C^2\Rd
	\left[
		\ah^2\alpha^2 \Da^2\Q 
		-
		2\alpha^2 \ah\eh\De\Da\Q
		+ 
		\alpha^2\eh^2\De^2\Q
		-
		\left(
			\frac{q\beta^{*}}{e^2}
			-
			\ah 
		\right)
		\alpha \eh\De\Q 
	\right]\ ,
\ee
\be\label{E:kkaQ}
\kk(\alpha \Q) = -\alpha\C^2 e^{-1}\sS
	\left[
		\alpha^2 \ah\De\Da\Q
		-
		\alpha^2 \eh\De^2\Q
		+
		\left(
			\frac{q\beta^{*}}{e^2}
			+
			\ah
		\right)
		\alpha\De\Q
	\right] \ ,
\ee
\be\label{E:kkQ}
\kk( \Q) = -\C^2 e^{-1}\sS
	\left[
		\alpha^2 \ah\De\Da\Q
		-
		\alpha^2 \eh\De^2\Q
		+
		\left(
			\frac{q\beta^{*}}{e^2}
			+
			0
		\right)
		\alpha \De\Q
	\right] \ ,
\ee
\be\label{E:LLaQ}
\LL(\alpha \Q) = \frac{\C}{\alpha}
	\left[ 
		\frac{2}{3}\tN\ah\Da(\alpha^2\Da\Q)
		+
		\left(
			\tE e \alpha \ah
			- 
			\frac{2}{3}\tN\alpha^2\eh 
		\right)
		\De\Da\Q
		-
		\tE q \beta \De^2\Q
		+
		\tE \frac{q}{e}
		\left(	6
			-
			5 \beta
		\right)
		\De\Q
	\right]\ ,
\ee
\be\label{E:LLQ}
\LL( \Q) = \C
	\left[ 
		\frac{2}{3}\tN\alpha^2\ah\Da^2\Q
		+
		\left( \tE e \alpha \ah
			-
			\frac{2}{3}\tN\alpha^2\eh 
		\right)
		\De\Da\Q
		-
		\tE q \beta \De^2\Q
		+
		\alpha
		\left(
			\tE q\beta^{*}
			+
			\frac{2}{3}\tN\eh
		\right)
		\De\Q
	\right] \ .
\ee
It is important to consider the magnitude of each of these operators to determine their significance in the equation of motion given in equation (\ref{E:eomcompact}).  Figures \ref{F:sinphi}, \ref{F:sindpom}, and \ref{F:damp} show the resulting magnitudes as a function of the test body eccentricity $e$ for the terms between the square brackets in line 1, 2, and 3 of equation (\ref{E:eomcompact}), respectively.

\subsection{Analysis of the Expansion Terms}

In this section we focus on the $\sin\phi$ part of equation (\ref{E:eomcompact}) -- the first line of the equation of motion for the resonance angle.  As written in this compact form, there are only five contributing terms contained within the square brackets.  The first term, $\N p n\planet \sin\phi$, when considered on its own, has been the focus of intense study in the circular restricted three body problem, and is called the pendulum model for obvious reasons. This model can be generalized by expanding each term within the square brackets via equations (\ref{E:auxcoeff}), (\ref{E:H}), (\ref{E:shorthand}), (\ref{E:JaQ}), and (\ref{E:JQ}) and keeping only the lowest eccentricity order contributions from each.  Doing so gives the result
\be
\ba
\ddot{\phi} \simeq -\sin{\phi} [ &-3/2 \ah \C (p n\planet + \dot{\phi}) \FI \ e^q
			-q^2\C  \FI \dot{\phi} \ e^{q-2}
			+2\C^2q\ah\FI (2\alpha \Da \fss + \fss)\ e^q\\
			&+\C^2q^2e\planet \fsS \FI \cos{\pomdiff} \ e^{q-3}
			-\C^2q^3(q-2)\FI^2\cos\phi \ e^{2(q-2)}] \ .
\ea
\ee
For non-circular perturber orbits, there is an additional contribution proportional to $e\planet^2$ that has been excluded here, which may be significant given large enough values of $e\planet$. Notice that, depending on the particular value of $q$, there are up to four different orders of eccentricity that show up, so we collect them in eccentricity order,
\be\label{E:EOMlowest}
\ba
\ddot{\phi} \simeq - \sin \phi \Big[ & \C \FI e^q \left\{ 2 q\ah\C \left(2\alpha \Da \fss + \fss\right) -3(p+q) p n\planet - \left(3(p+q) + q^2 e^{-2} \right) \dot{\phi} \right\} \\
					&\quad \quad +  \C^2 q^2 e^{q-3} \FI \left\{ e\planet \fsS \cos \pomdiff - q(q-2) e^{q-1} \FI \cos \phi \right\} \Big] \ .
\ea
\ee
However, the number of terms shown here can be further reduced.   Although we are interested in a lowest eccentricity order study, we must also consider the lowest order terms in $\C \propto \mu$ as well.  For a 2:1 period ratio, test body mean motion $n = 2 \pi$, mass ratio $\mu \sim 10^{-3}$, and perturber eccentricity $e\planet = 0.1$, Figure \ref{F:sinphi} depicts the corresponding relative strengths of each term contributing to the $\sin \phi$ part (first line) of equation (\ref{E:eomcompact}). 

For eccentricities $e \sim 0.1$, two terms in the expansion dominate. One term is the usual pendulum term, but the other term is something like a modified damping term (i.e., a term multiplying the time derivative $\dot{\phi}$).  Although, $|\dot{\phi}|$ is generally less than unity for instances of libration, contributions from the $\dot{\phi}$ term can become quite large for sufficiently small eccentricities due to its $e^{-1}$ dependence.  In the figure, contributions from $\dot{\phi}$ and the pendulum term have the same sign (both curves are dashed lines, meaning less than zero on the log-log plot). To achieve nodding by way of the $\dot{\phi}$ term, one requires $\dot{\phi} < 0$ \emph{in addition to} satisfying the constraint on the upper bound of the eccentricity, $e \le 0.1$.  The condition that $\dot{\phi} < 0$ for nodding to occur provides a simple explanation for the observation that the resonance angle tends to circulate on average over  secular timescales in a preferred direction, usually resulting in a graph of $\phi$ versus $t$ which is reminiscent of a stair case.  The figure actually shows that this term is comparable to the pendulum term for all values of eccentricity, however it works to counteract the pendulum term for instances where $\dot{\phi} > 0$, and has opportunity to overwhelm the pendulum term only for $|\dot{\phi}| \gtrsim 2$.

Another term of significance appears in equation (\ref{E:EOMlowest}). For sufficiently small test body eccentricities, the term that is proportional to $e^{-2}\cos\phi$ plays an important role. As shown in Figure \ref{F:sinphi}, the $\cos \phi$ coefficient (red curve) surpasses the pendulum term (cyan curve)  for eccentricities lower than $\sim 0.03$. The greatest lower bound on eccentricity required for this term to take effect is
\[
e \lesssim \left(\frac{\mu\FI\sqrt{\alpha}}{6}\right)^{1/3} \ .
\]
When evaluated using the typical value $\mu = 10^{-3}$ in the external 2:1 case ($\FI \sim 0.27$), the above expression gives the restriction $e \lesssim 0.033$. Note that the $e^{-2} \cos \phi$ term is not present for those resonances with $q=2$ (e.g., 3:1, 5:3, 7:5, etc...), and has opposite sign for resonances with $q=3$ (e.g., 4:1, 5:2, 7:4, etc...) and above.  However, for resonances of higher rank (higher $q$ values), the eccentricity order increases relative to the order of the pendulum and the $\dot{\phi}$ terms, diminishing its importance.

Keeping only the two largest terms from Figure \ref{F:sinphi}, the equation of motion can be reduced to the form
\be\label{E:REDcos}
\ddot{\phi} \approx \C \FI \sin{\phi} \left[  6n\planet e + e^{-1} \dot{\phi} - \C\FI e^{-2}\cos\phi \right]\ .
\ee
This equation exhibits some of the nodding behaviors we see in the full problem (see section \ref{S:full}).  The parameters and initial values must be fine-tuned for cases where eccentricities (and periastra) are independent of time (i.e., only special values allow for nodding). Under conditions where the eccentricity has time dependence, however, nodding is a robust phenomenon(it is much easier to find parameters for which nodding occurs).  For the sake of simplicity, we adopt the parametric description for the time dependence of eccentricity
\be\label{E:et}
e(t) = e_0 - e_a \sin^2(\omega_e t) \ .
\ee
In many instances of nodding found from the numerical studies in section 2 but not featured in the figures, the test body's orbital eccentricity evolves through large swooping double arches, spanning several orders of magnitude, reaching values as high as a few times $10^{-1}$ and as low as $10^{-2} - 10^{-3}$ (see top panel of Figure \ref{F:Ext_nod}). This double arched pattern exhibited by the test particle's orbital eccentricity is governed by the secular time scale, which typically falls in the range $\sim 10^2 - 10^3$ libration times, but can be much longer (e.g., Michtchenko et al. 2008b). The ansatz of equation (\ref{E:et}) is used in order to model this behavior. 

\begin{figure}[htbp]
\begin{center}
\includegraphics[width=150mm]{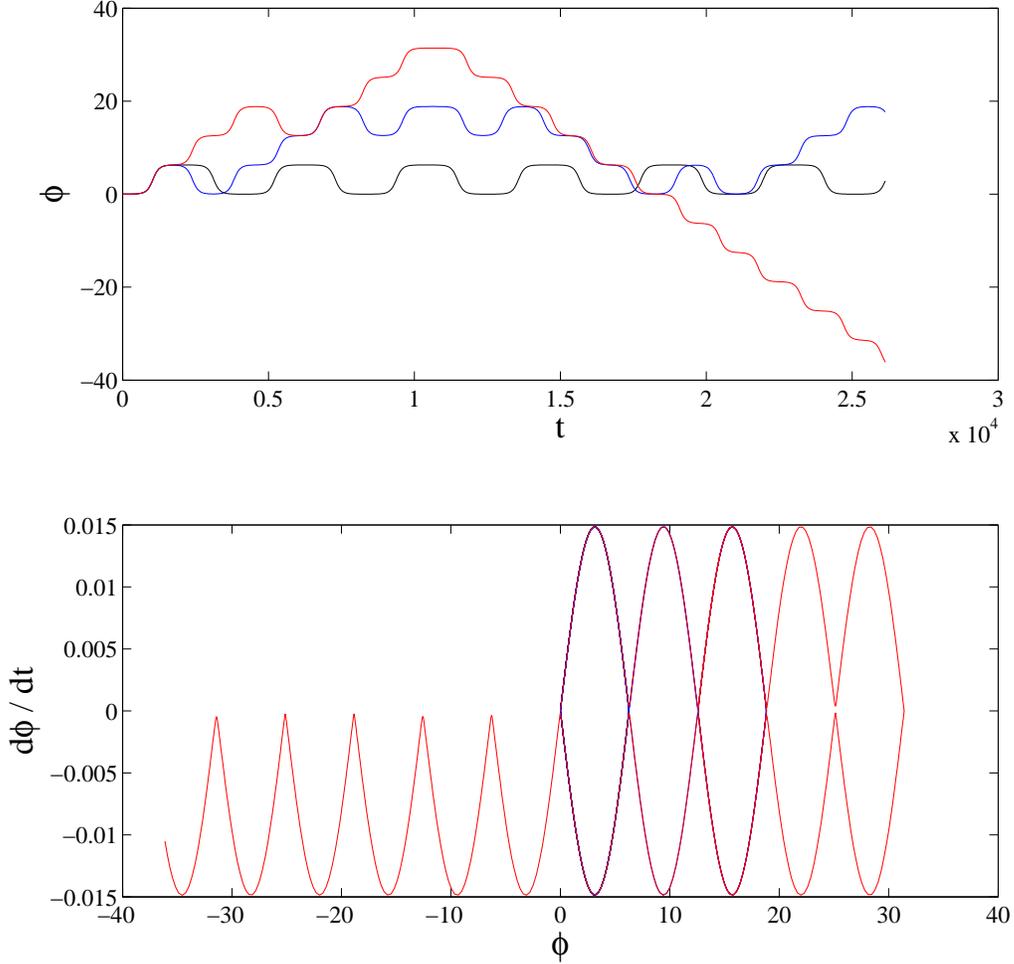}
\caption{Solutions to the pendulum equation with constant eccentricity (black curve) and with eccentricity that varies by 0.1\% on a secular timescale (blue and red curves).  The system is set in a dynamically sensitive state near a separatrix with $\phi = 0.001$ and $\dot{\phi} = 0$ and integrated using $4^{\text{th}}$ order Runge-Kutta scheme for 16 full secular cycles.  The eccentricity is initially small ($e = 0.001$), and the pendulum term has no knowledge of the inner body's eccentricity.  The introduction of time varying eccentricity (even with small amplitude) induces circulation in the motion of the pendulum.  The eccentricity cycle is phase shifted by $\pi$ between the red and blue curves, where the blue curve begins at peak eccentricity. The bottom panel shows the phase trajectory for the resonance angle solutions in the top panel.  The resonance angle is on phase trajectory that is very near a separatrix of the phase space. Note that in the phase diagram, the black curve is confined to $0 \lesssim \phi \lesssim 2 \pi$, the blue curve is confined to $0 \lesssim \phi \lesssim 6 \pi$, and the red curve is confined to $-12 \pi \lesssim \phi \lesssim 10 \pi$.  }
\label{F:pend_nod}
\end{center}
\end{figure}

\begin{figure}[htbp]
\begin{center}
\includegraphics[width=150mm]{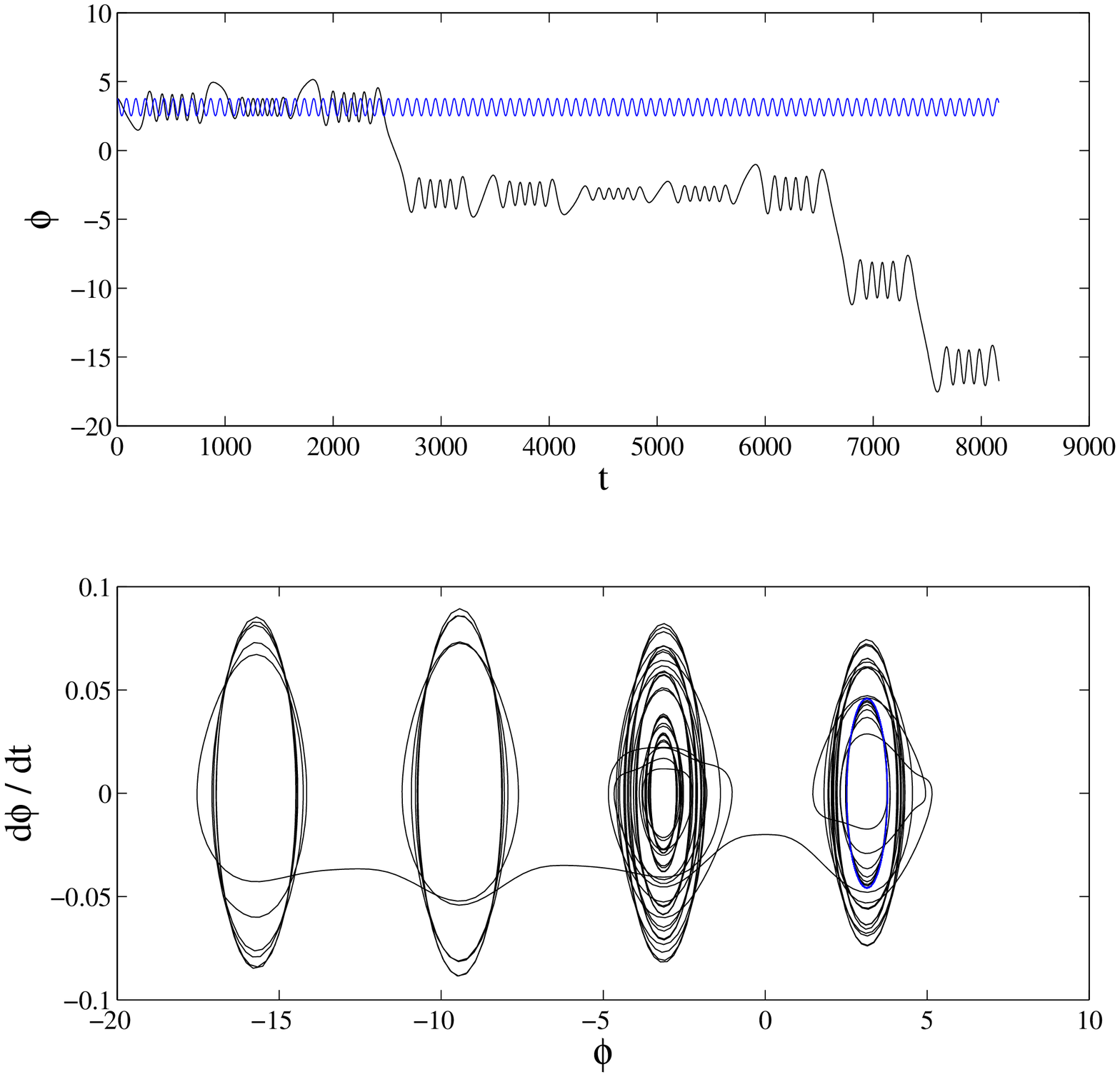}
\caption{Solutions to the pendulum equation for initial values $e = 0.1$, $\phi = 1.2 \ \pi$, and $\dot{\phi} = 0$. Both solutions presented here are for the pendulum term with (\emph{i}) constant eccentricity (blue curve) and (\emph{ii}) with eccentricity that varies by 99\% on a secular timescale (black curve).  When eccentricity is constant, the system remains in an oscillatory state with amplitude $\Delta \phi \approx \pi/4$ around the point$\phi \simeq \pi$. For time varying eccentricity, the librations of the resonance angle become unbound before 2 complete secular cycles, and nodding ensues. The bottom panel shows the phase trajectory for the resonance angle solutions in the top panel.  Oscillations of the resonance angle occur around points that are equivalently located at $\phi = \pi$ (modulo $2\pi$).}
\label{F:pend_nod2}
\end{center}
\end{figure}

\begin{figure}[htbp]
\begin{center}
\includegraphics[width=150mm]{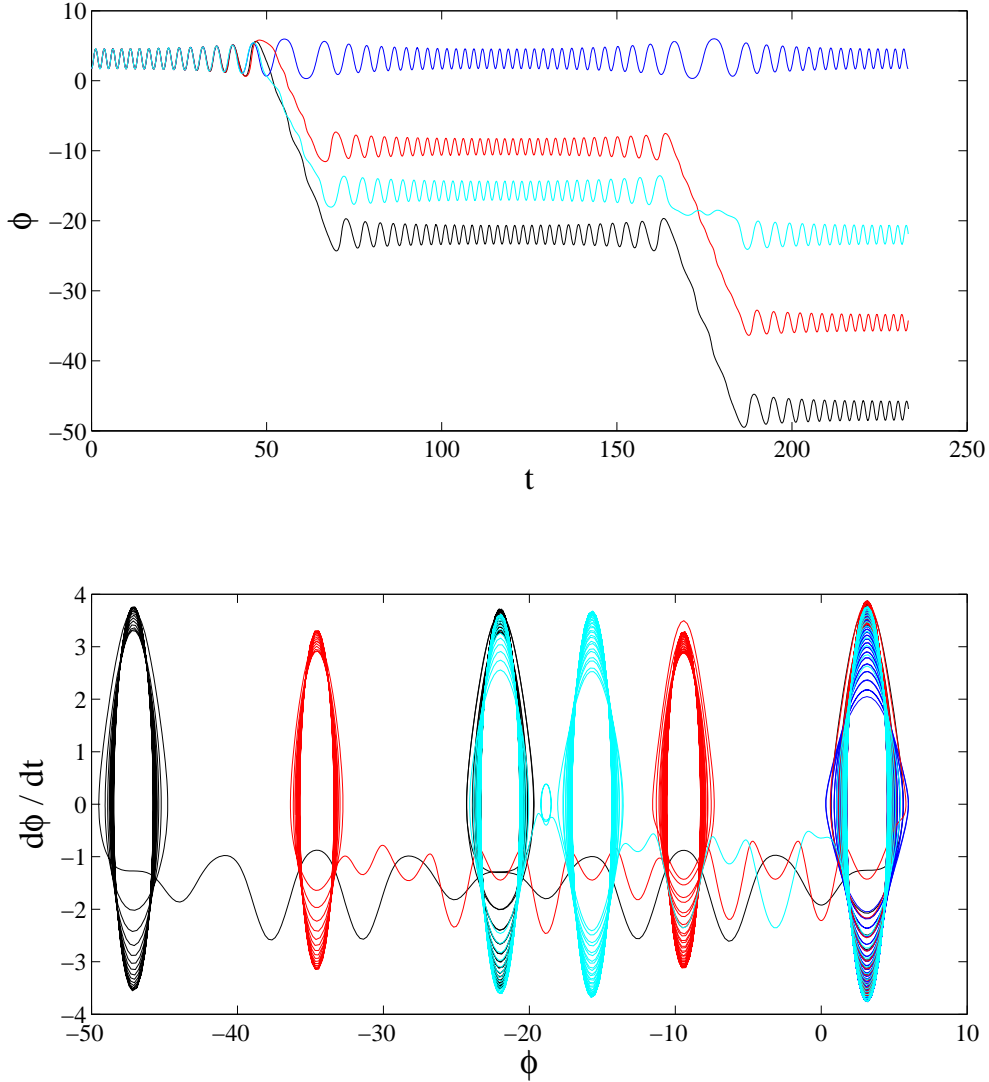}
\caption{The result of numerically integrating equation (\ref{E:REDcos}) using parameters $T_{orb} = 100$ days, $T_{sec} = T_{orb}\times10^{2.94}$, $\mu = 10^{-3}$, $e_0 = 0.32$, $e_a = 0.28$, $\FI = 1$, and the orbiting bodies in a perfect 2:1 orbital period ratio. The blue curve shows the solution for the pendulum term alone; the black curve shows the solution for the pendulum term plus the term that goes like $\dot{\phi}$; the cyan curve includes the pendulum term and the $e^{-2}\cos\phi$ term; the red curve shows the solution when including all terms in equation (\ref{E:REDcos}). The only model out of the four depicted here that doesn't exhibit nodding is that consisting of the pendulum term in isolation. Inclusion of the two terms of lower eccentricity order and in any combination results in nodding. The timescale depicted in the top panel is in years. The bottom panel shows the phase trajectory for the resonance angle solutions in the top panel.}
\label{F:SIMeomA}
\end{center}
\end{figure}

To demonstrate that this model exhibits some nodding behaviors, we integrate equation (\ref{E:REDcos}) using an adaptive fourth order Runge-Kutta integration scheme with $e=\text{constant}$ and with $e(t)$ given by equation (\ref{E:et}). We note that this is a simple study in which the eccentricity evolution being used here is totally prescribed without feedback from the specific orbital parameters, so we do not expect the model to exhibit all of the intricate details exhibited by the full 3-body simulations presented in section \ref{S:full}. For the sake of definiteness, we take $n\sim1$, $\C \sim 10^{-2}$, and $\omega_e \sim 10^{-3}$ yr$^{-1}$ -- the precise values of these parameters will depend on the orbital angles and mean motions, and the values we choose corresponds to orbital periods on the order of years, not days. Under the parametrically evolving eccentricity of equation (\ref{E:et}), nodding is a robust phenomenon, even when considering the pendulum term alone. Guided by the pendulum model of the circular restricted 3-body problem, we take the initial conditions $(\phi,\dot{\phi}) = (\epsilon, 0)$, which places the system in a dynamically vulnerable position near a separatrix  -- any perturbations that supply additional action should eventually cause the pendulum to circulate rather than oscillate. Figure \ref{F:pend_nod} demonstrates how a small amount of eccentricity variation can send such a system into bouts of nodding. However, the system need not be prepared in such a dynamically sensitive manner to see nodding. Figure \ref{F:pend_nod2} shows a system that would be stable in the absence of variable eccentricity, but where the inclusion of sufficient eccentricity cycling provides enough added action to induce nodding. In both Figures \ref{F:pend_nod} and \ref{F:pend_nod2}, we use $\FI \simeq 0.27$, which is the approximate value obtained from equation (\ref{E:shorthand}) corresponding to a 2:1 period ratio with $\alpha \simeq 0.63$ (see MD99). Figure \ref{F:SIMeomA} shows the results when taking $\FI = 1$ and using various combinations of the terms in equation (\ref{E:REDcos}) along with a time varying eccentricity as defined in (\ref{E:et}). This figure demonstrates that the pendulum model alone is not enough to recover nodding behaviors, but nodding does appear in models that combine the pendulum term with either of the two additional terms in equation (\ref{E:REDcos}). Taken together, Figures \ref{F:pend_nod} -- \ref{F:SIMeomA} show that nodding behavior arises naturally in modified pendulum equations, such as those resulting from the expansion of the previous section.

It may be worth noting that another term in equation (\ref{E:EOMlowest}) is of order $e^{-2}$, which was ignored due to its contribution from $e\planet$.  However, this term could be of importance in cases where the test particle's eccentricity is very low ($e \ll e\planet$). Including this term leaves us with a form dependent upon $\pomdiff$, 
\be
\ddot{\phi} \approx \C \FI \sin{\phi} \left[  6n\planet e + e^{-1} \dot{\phi} - \C\FI e^{-2}\cos\phi - \C \fsS e\planet e^{-2}\cos \pomdiff \right]\ .
\ee
This equation will exhibit nodding, even for constant eccentricity (no time variation), as long as the eccentricity is sufficiently low and the periastron circulates on secular timescales.  

We caution that this model certainly does not exhaust all possible terms that can lead to resonance angle nodding.  In the regime where the perturbing planet's eccentricity is non-negligible $e_p \gtrsim 0.1$, terms of first order in the planet's eccentricity can become significant.  In particular, the term in the disturbing function that goes like $\sim e_p \cos(\phi + \pomdiff)$ can lead to some interesting modulations in the resonance angle, which ultimately may drive the resonance angle's phase into the vicinity of a separatrix and induce nodding (K.~Batygin, private communication).

The model developed in this section contains one important deficiency.  The inner separatrix of the external resonance problem is completely missing from this model along with the unique characteristics distinguishing the external resonance from the internal one.  This separatrix originates because of a bifurcation that occurs for the external resonance when $e \sim 0.1$, and is responsible for asymmetric resonances that have been observed in simulations for that regime (e.g., Callegari et al. 2004, Lee 2004). As a result, $\phi = \pi$ (or a point nearby) becomes a hyperbolic fixed point.  Given that the test body's eccentricity must be sufficiently large before this bifurcation occurs may suggest that more terms of the disturbing function are required in the time averaged treatment to recover these dynamics.  More work must be done in order to elucidate this issue.

\section{Conclusion}

This paper presents an investigation of nodding behavior for planetary
systems that are near mean motion resonance, with a focus on resonance
angles corresponding to the 2:1 MMR. For systems that experience the
nodding phenomenon, the resonance angle librates (oscillates) for
several cycles, then circulates for one or more cycles, and then
resumes its oscillatory motion (libration). The process repeats, so
that even though the resonance angle is primarily in oscillation, the
phase of the resonance angle nonetheless accumulates over time. In the
extreme version of this behavior, the resonance angle can oscillate
for one cycle, circulate for the next cycle, and then repeat the
process; the resonance angle moves continually back and forth between
the two types of motion, so that the resonance angle has an effective
period of oscillation that is $\sim2$ times longer than the usual
period for MMR (see Figures 3, 8, and 9). Our numerical exploration 
of Section 2 shows that the nodding phenomenon arises in a wide 
variety of planetary systems (see Figures 1 -- 9). 

Nodding can be described as complex motion near a separatrix in the
phase space of the resonance angle.  Both internal and external
resonances can exhibit nodding, but there exist prominent qualitative
differences in the nodding signatures between the two configurations.
The phase space for internal resonances contain one separatrix,
whereas the phase space for external resonances can contain two
distinct separatrices. The qualitative differences are mainly due to
the existence of asymmetric external resonances, which arise when the
orbital eccentricity for the outer (smaller) body becomes sufficiently
large. Circulation of the resonance angle over secular times is common
when the planet's orbital eccentricity is sufficiently large, i.e.,
$e\planet \gtrsim$ 0.02. Nonetheless, the essential ingredients for 
nodding behavior are the presence of a separatrix and a means to 
cross it. Note that separatrix crossing can occur in sufficiently complex systems 
(e.g., the driven, inverted pendulum; Acheson 1995) and can arise due 
to chaos (e.g., Morbidelli \& Moons 1993).

Exoplanets that transit their host stars are expected to exhibit
transit timing variations when additional perturbing bodies are
present, and the variations are greatest when the perturbing body is
either in or near mean motion resonance with the transiting body. For
systems where the perturbing body is near MMR, there is a possibility
that the resonance angle may undergo nodding. In such systems, both
the amplitude and the period of the TTVs depend on the window of time
over which the system is observed (see Figures \ref{F:ttv} and
\ref{F:ttv2}). If the observations are made over a time interval where
the resonance angle of the system oscillates for many cycles, the TTVs
have their usual interpretation. In the limit where the resonance
angle goes back and forth between oscillating and circulating every
cycle, the effective frequency of the TTVs is lower (than in the
purely oscillatory case) by a factor of $\sim2$. Nodding behavior that
is intermediate between these two cases is also possible, and produces
TTVs with intermediate frequencies (see the power spectra in the
bottom panels of Figures \ref{F:ttv} and \ref{F:ttv2}). Note that the
amplitudes of the TTVs vary with the mode of nodding/oscillation (for
the same masses). The added complexity in the dynamics due to nodding
can thus introduce corresponding difficulties in interpreting the
source of transit timing variations. This possible complexity should
be kept in mind when searching for hidden exoplanets through
measurements of transit times.

Using Lagrange's planetary equations of motion, we have derived a set
of generalized equations of motion for the resonance angle by
including additional terms in the expansion in order to account for
nodding behavior (see Section \ref{S:eom}).  This derivation uses the
time-averaged disturbing function and initially keeps terms up to
second order in eccentricity. As expected, the initial expansion
includes a large number of terms.  We then performed an analysis that
uses results from the full numerical treatment to determine the
relative sizes of the various terms in the expansion over the
parameter space of interest (see Figures 10, 11, and 12). For some
parameter values, we found that some of the ``higher order'' terms
(those left out of the MD99 derivation) can dominate over those terms
commonly used in deriving the pendulum model, which provides a
standard description for MMR.

Given the expected magnitudes of the expansion terms, we have
constructed a modified model for MMR, where the equation of motion
includes two additional terms and thus allows for more complex
dynamics (see equation [\ref{E:REDcos}]).  This equation exhibits the
nodding behavior found for interior resonances of the full
problem. However, nodding only occurs for particular values of the
eccentricity and argument of periastron, where both variables are
considered as constants. We then generalized this model one step
further by allowing the eccentricity to change with time (see equation
[\ref{E:et}]), motivated by the secular cycles of eccentricity
variation often seen in two planet systems (including the orbits of
Jupiter and Saturn in our solar system -- see MD99). With this
generalization to include eccentricity cycles, nodding is a robust
phenomenon and occurs for a wide range of the other parameters. Even
when considering the pendulum term in isolation, the cycling of
eccentricity on secular timescales can provide sufficient
perturbations to induce nodding behaviors in certain dynamically
vulnerable configurations.

The model from section \ref{S:eom} contains at least one deficiency.
The inner separatrix of the external resonance problem is
missing from the model, along with the unique characteristics
distinguishing the external resonance from the internal one.  This
additional separatrix originates for the external resonance for larger 
outer body eccentricities, when $e \sim 0.1$, and is reminiscent of the
pitchfork bifurcation shown in Figure \ref{F:dyn_map}.  This separatrix
is responsible for the asymmetric resonances that
have been observed for that regime.  As a result, some region in the
neighborhood of $\phi = \pi$ contains a hyperbolic fixed point.  Given
that the eccentricity of the test body must be significantly large
before this bifurcation occurs, additional terms in the expansion of
the disturbing function may be required to recover these
dynamics. This issue is left for future work.

The results of this work, and the existence of nodding phenomena in
general, have two principal implications: [1] Planetary systems near
MMR display complex and sometimes unexpected behavior, so that nodding
poses a rich set of dynamical questions for further work. It would be
interesting to determine the minimal requirements for a dynamical
system to exhibit nodding, and to explore the relationship between
nodding and chaotic motion. [2] The main application of this work to
observations lies in the interpretation of transit timing variations.
If an observed system with TTVs experiences nodding, then the inferred
mass and orbital elements of the unseen perturbing body could vary,
depending on the time interval of observation (Figures \ref{F:ttv} and
\ref{F:ttv2}). A more complete exploration of parameter space, with a
focus on the TTV signals (analogous to Veras et al. 2011), should be
undertaken in the future.

\acknowledgments

This paper benefited from discussions with many colleagues, especially
Konstantin Batygin, Philip Holmes, Greg Laughlin, Man Hoi Lee, and
Matthew Payne. We also thank an anonymous referee for useful comments
that improved the manuscript.  This work was supported by NSF grant
DMS-0806756 from the Division of Applied Mathematics, a Rackham
Predoctoral Fellowship from the University of Michigan (JAK), NASA
grant NNX11AK87G (FCA), and NSF grants DMS-0907949 and DMS-1207693 (AMB).

\end{document}